\documentclass[fleqn,usenatbib]{mnras}
\pdfoutput=1
\usepackage{newtxtext,newtxmath}

\usepackage[T1]{fontenc}
\usepackage{ae,aecompl}

\usepackage{graphicx}	
\usepackage{amsmath}	
\usepackage{makecell}
\graphicspath{
{./Figures/}
}

\newcommand{\f}{\frac}
\newcommand{\tx}{\text}
\newcommand{\Li}{\mathcal{L}}

\title[2D Temperature Models for Analysing Phase Curves]{Testing 2D temperature models in Bayesian retrievals of atmospheric properties from hot Jupiter phase curves}

\author[J. Yang et al.]{
Jingxuan Yang,$^{1}$\thanks{E-mail: jingxuan.yang@hertford.ox.ac.uk}
Patrick G.J. Irwin,$^{1}$
Joanna K. Barstow$^{2}$
\\
$^{1}$Department of Physics, University of Oxford, Parks Road, Oxford OX1 3PU, UK\\
$^{2}$School of Physical Sciences, The Open University, Walton Hall, Milton Keynes, MK7 6AA, UK\\
}

\date{Accepted XXX. Received YYY; in original form ZZZ}

\pubyear{2023}

\begin{document}
\label{firstpage}
\pagerange{\pageref{firstpage}--\pageref{lastpage}}
\maketitle

\begin{abstract}
Spectroscopic phase curves of transiting hot Jupiters are spectral measurements at multiple orbital phases, giving a set of disc-averaged spectra that probe multiple hemispheres. 
By fitting model phase curves to observations, we can constrain the atmospheric properties of hot Jupiters,  such as molecular abundance, aerosol distribution, and thermal structure, which offer insights into their atmospheric dynamics, chemistry, and formation.
We propose a novel 2D temperature parameterisation consisting of a dayside and a nightside to retrieve information from near-infrared phase curves and apply the method to phase curves of WASP-43b observed by \textit{HST}/WFC3 and \textit{Spitzer}/IRAC.
In our scheme, the temperature is constant on isobars on the nightside and varies with cos$^n$(longitude/$\epsilon$) on isobars on the dayside, where $n$ and $\epsilon$ are free parameters.
We fit all orbital phases simultaneously using the radiative transfer package \textsc{NEMESISPY} coupled to a Bayesian inference code. 
We first validate the performance of our retrieval scheme with synthetic phase curves generated from a GCM and find that our 2D scheme can accurately retrieve the latitudinally averaged thermal structure and constrain the abundance of H$_2$O and CH$_4$. 
We then apply our 2D scheme to the observed phase curves of WASP-43b and find: (1) the dayside temperature-pressure profiles do not vary strongly with longitude and are non-inverted; (2) the retrieved nightside temperatures are extremely low, suggesting significant nightside cloud coverage; (3) the H$_2$O volume mixing ratio is constrained to $5.6\times10^{-5}$--$4.0\times10^{-4}$, and we retrieve an upper bound for CH$_4$ mixing ratio at $\sim$10$^{-6}$. 
\end{abstract}


\section{Introduction}
Exoplanet surveys suggest planets are common around stars in our galaxy \citep{winn_occurrence_2015}. The diversity in their characteristics from system architecture to bulk properties poses challenging questions in the theory of planetary formation \citep{mordasini_extrasolar_2009}. Gaseous giant planets with close-in orbits (period $<$ 10 days), dubbed hot Jupiters, are a key piece of the puzzle for two reasons: (1) they likely undergo significant orbital migration and play an important role in shaping planetary system architecture \citep{dawson_origins_2018}, and (2) they are the easiest targets for spectroscopic characterisation, and the constraints on their atmospheric properties give valuable insights into planetary formation \citep{madhusudhan_atmospheric_2017, mordasini_imprint_2016}. 

The spectral appearance of hot Jupiters is determined by the opacity structure and the thermal structure of their atmospheres. 
Conversely, by fitting spectra generated from atmospheric models to observations, we could constrain the atmospheric properties of these planets in a process known as atmospheric retrievals \citep{irwin_nemesis_2008, madhusudhan_temperature_2009, line_systematic_2013, changeat_taurex3_2020, cubillos_pyrat_2021, macdonald_catalog_2023}. Two observing methods are widely used: (1) transmission spectroscopy, which measures the stellar light filtered through the planetary limb during primary transits \citep{barstow_consistent_2017, sing_continuum_2016}; (2) eclipse spectroscopy, which extracts the dayside emission spectra by monitoring the combined stellar and planetary flux during secondary eclipses \citep{lee_optimal_2012, mansfield_unique_2021}.  Such observations have been done at both high resolution \citep[e.g.,][]{brogi_retrieving_2019} from ground-based facilities and at low-resolution using space telescopes \citep[e.g.,][]{wakeford_complete_2017}, resulting in a myriad of atomic and molecular detections (e.g., Fe: \citealt{hoeijmakers_atomic_2018}; Na: \citealt{snellen_ground-based_2008}; H$_2$O: \citealt{evans_detection_2016}; CO$_2$:  \citealt{the_jwst_transiting_exoplanet_community_early_release_science_team_identification_2022}).  

A major challenge in the analysis of low-resolution emission spectra is the degeneracy between thermal structure and molecular abundance; in particular, modelling highly inhomogeneous thermal structure with a single temperature-pressure (TP) profile can lead to significant biases in retrieved molecular abundance \citep{feng_impact_2016, blecic_implications_2017, taylor_understanding_2020}. 
This degeneracy can be partially broken by measuring the emission spectra of transiting hot Jupiters at multiple orbital phases. 
As hot Jupiters are likely tidally locked to their stars due to the short orbital separations, it is straightforward to relate the orbital phases to the central longitudes of the visible hemisphere. 
Note that we assume the planets have edge-on orbits, and we define the equator to be in the orbital plane. Such observations, called `phase curves,' allow us to constrain the thermal structure of the atmospheres better, resulting in better constraints on the chemical abundance as well. 

In order to retrieve information from a set of phase curves using Bayesian inference, we first need to construct appropriate parametric atmospheric models. If we want to analyse all orbital phases simultaneously, then the models must describe the entire observable atmosphere to generate disc-averaged emission spectra at multiple orbital phases. Crucially, we need multidimensional temperature models that can capture the longitudinal variation of thermal structure in the pressure ranges probed by emission spectroscopy. 
The model should also contain as few parameters as possible to ensure Bayesian parameter estimation can be done in reasonable time, and to avoid overfitting.
There are two main approaches to this problem. The first approach is to split the atmosphere into disjoint regions, where the thermal structure in each region is modelled with a one-dimensional TP model. 
For example, \cite{feng_2d_2020} split the atmosphere into a dayside and a nightside, 
whereas \cite{changeat_exploration_2021} further model an additional hot spot within the dayside, and \cite{irwin_25d_2020} divide the atmosphere into meridian bands with linearly interpolated temperature maps between the bands. 
The second approach is to construct highly-simplified three-dimensional analytical atmospheric models.  
For example, \cite{dobbs-dixon_gcm-motivated_2022} propose a 3D model by separating radiative and convective components from Global Circulation Model (GCM) outputs, whereas \cite{chubb_exoplanet_2022} prescribe a 3D model by restricting heat transfer to diffusion and zonal winds. 

The studies summarised above offer unique and valuable perspectives on the analysis of hot Jupiter phase curves. However, it is difficult to compare the different retrieval schemes for several reasons. Firstly, the ways in which the retrieval schemes are validated differ significantly across studies. 
The most direct way to assess the performance of a retrieval scheme is first to create synthetic data from a model atmosphere, then test how well the retrieval scheme can recover the input atmospheric properties. 
Out of the studies that include such validation tests, the synthetic data are often generated from toy models resembling the temperature parameterisations of the retrieval schemes, so it needs to be clarified how well the retrieval models can perform on data generated from more realistic atmospheric models.
Secondly, there are multiple modelling steps within each retrieval study, for example, the modelling of thermal structure, the modelling of chemical abundance, and the modelling of radiative transfer, which can all vary across studies. 
Thirdly, the studies that perform analysis of real observations often do not retrieve on exactly the same data set, which hinders the comparison of the retrieved constraints. 

In this work, we propose a novel 2D retrieval scheme (model 4 in section \ref{sec:simple_T_models}) that can be used to retrieve chemical abundance and thermal structure from hot Jupiter phase curves.
In this model, the temperature is a function of pressure and longitude.
The model is split into a dayside and a nightside: on the dayside, the temperature varies with cos$^n$(longitude/$\epsilon$) on isobars, where $n$ and $\epsilon$ are free parameters, and on the nightside the temperature is constant on isobars.
We use this scheme, together with several other simpler 2D retrieval schemes for comparison, to retrieve molecular abundance and latitudinally averaged thermal structure from phase curves of WASP-43b observed by \textit{Hubble Space Telescope}/Wide Field Camera 3 (\textit{HST}/WFC3) and \textit{Spitzer}/Infra-Red Array Camera (\textit{Spitzer}/IRAC).  
We first validate the performance of the 2D schemes by retrieving atmospheric properties from synthetic phase curves generated from a GCM-based model of WASP-43b, where the `ground truth' is known, so we can assess the accuracy of the retrieved properties. 
We then apply the 2D schemes to the observed \textit{HST}/WFC3 and \textit{Spitzer}/IRAC phase curves of WASP-43b. 
We also compare the 2D approach to the phase-by-phase approach, where the spectrum at each orbital phase is analysed separately, in appendix \ref{app:phase_by_phase}. 

This paper is structured as follows. 
In section \ref{sec:methodology}, we describe our routine for simulating spectroscopic phase curves, our 2D temperature models, and our retrieval set-up. 
Section \ref{sec:preliminary} demonstrates that simplified atmospheric models can reproduce the synthetic data generated from a GCM.
Section \ref{sec:gcm_test} presents the retrieval results on synthetic GCM phase curves using our 2D temperature models, followed by application to the observed phase curves in section \ref{sec:observation_test}. 
We discuss the implications of our results and compare them with past studies in section \ref{sec:discussions} and end with a conclusion in section \ref{sec:conclusions}.

\section{Methodology}
\label{sec:methodology}

This work aims to assess the performance of a novel 2D parametric temperature model (model 4) in retrieving atmospheric properties from low-resolution spectroscopic phase curves. 
We also test three simpler models (model 1, 2, and 3) for comparison.
The data we model are the phase curves of WASP-43b, as presented in \cite{stevenson_spitzer_2017}, and synthetic phase curves of the same resolution simulated from a GCM-based model. 
The planet WASP-43b is a hot Jupiter around a K7 star discovered by \cite{hellier_wasp-43b_2011}, with planetary parameters of 2.034$\pm$0.052 Jupiter mass and 1.036$\pm$0.019  Jupiter radii as given by \cite{gillon_trappist_2012}. Due to its short 19.5-hour orbit and large planet-to-star flux ratio, WASP-43b is a prime target for phase-resolved spectroscopic observations.
The observation contains 15 phase curves from the \textit{HST}/WFC3 instrument, which are binned in equally spaced bins of width 0.035 $\upmu$m spanning the wavelength range 1.1425-1.6325 $\upmu$m, and two phase curves from \textit{Spitzer}/IRAC broad channels centred at 3.6 and 4.5 $\upmu$m. 

In this section, we describe the GCM used for simulating synthetic phase curves in \ref{sec:gcm_intro}, our procedure for the radiative transfer calculation in \ref{sec:radiative_transfer}, and our method for calculating disc-averaged spectra in \ref{sec:disc-average}. 
We describe our 2D atmospheric temperature models in \ref {sec:simple_T_models}, our retrieval set-up in \ref{sec:retrieval_set_up}, and our Bayesian inference scheme in \ref{sec:nested_sampling}.

\subsection{GCM data}
\label{sec:gcm_intro}
    
\begin{figure}
\includegraphics[width=\columnwidth]{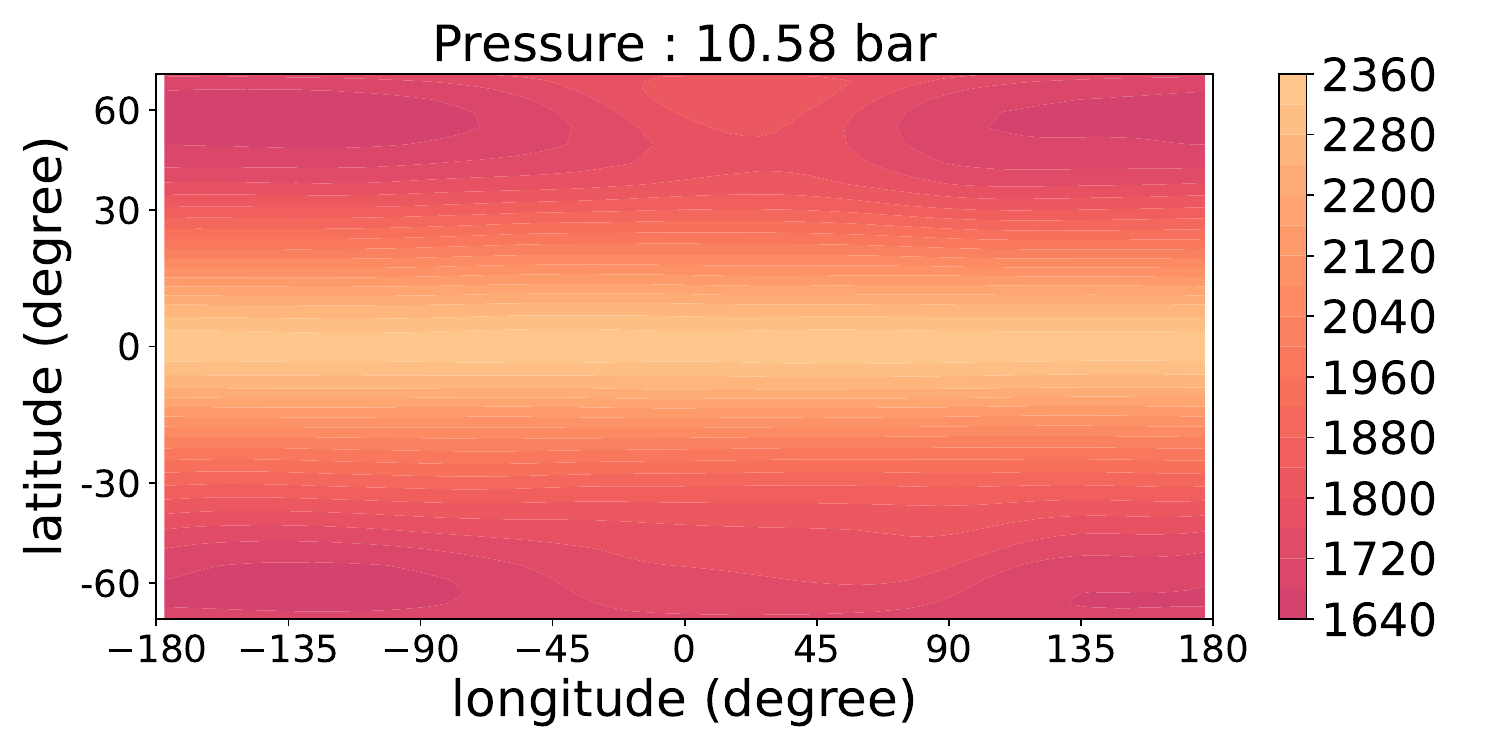}
\includegraphics[width=\columnwidth]{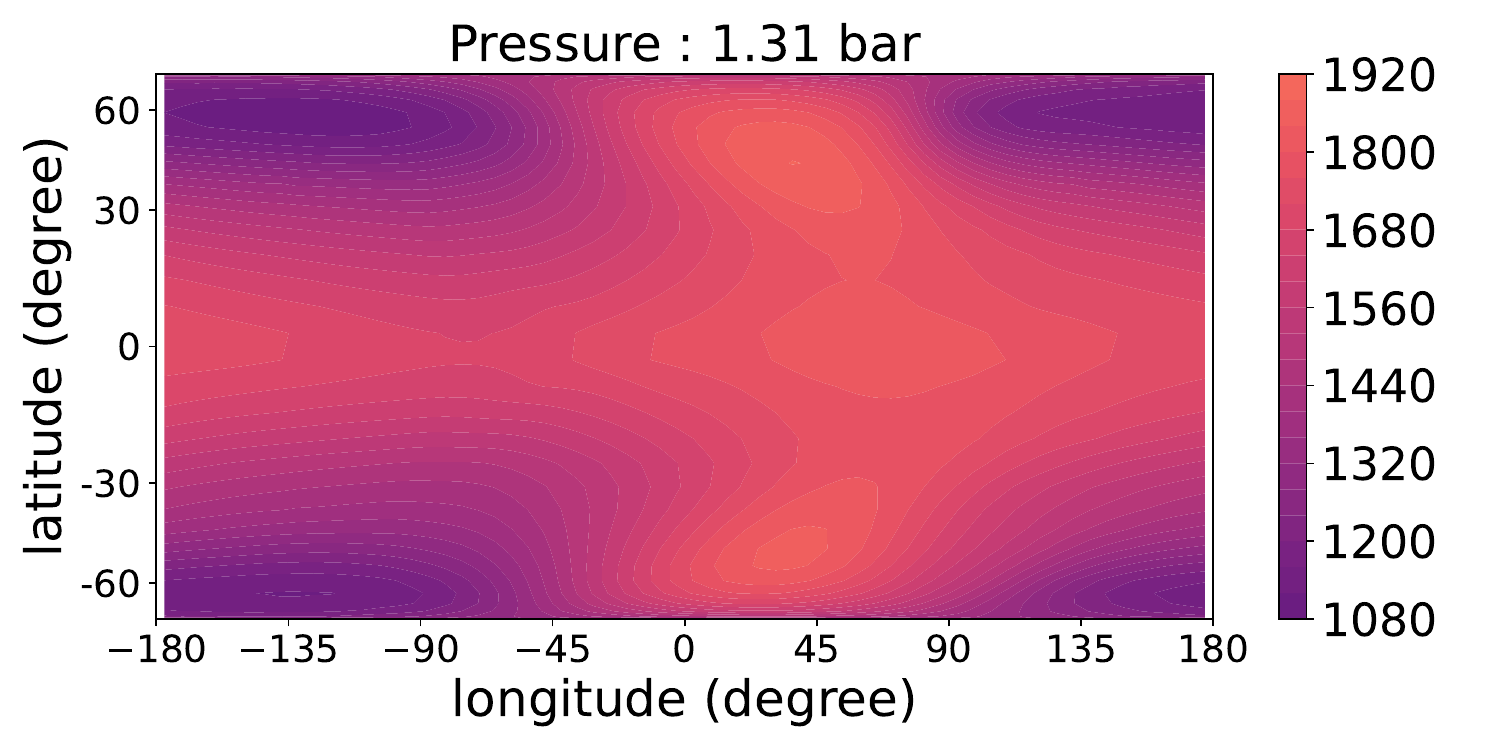}
\includegraphics[width=\columnwidth]{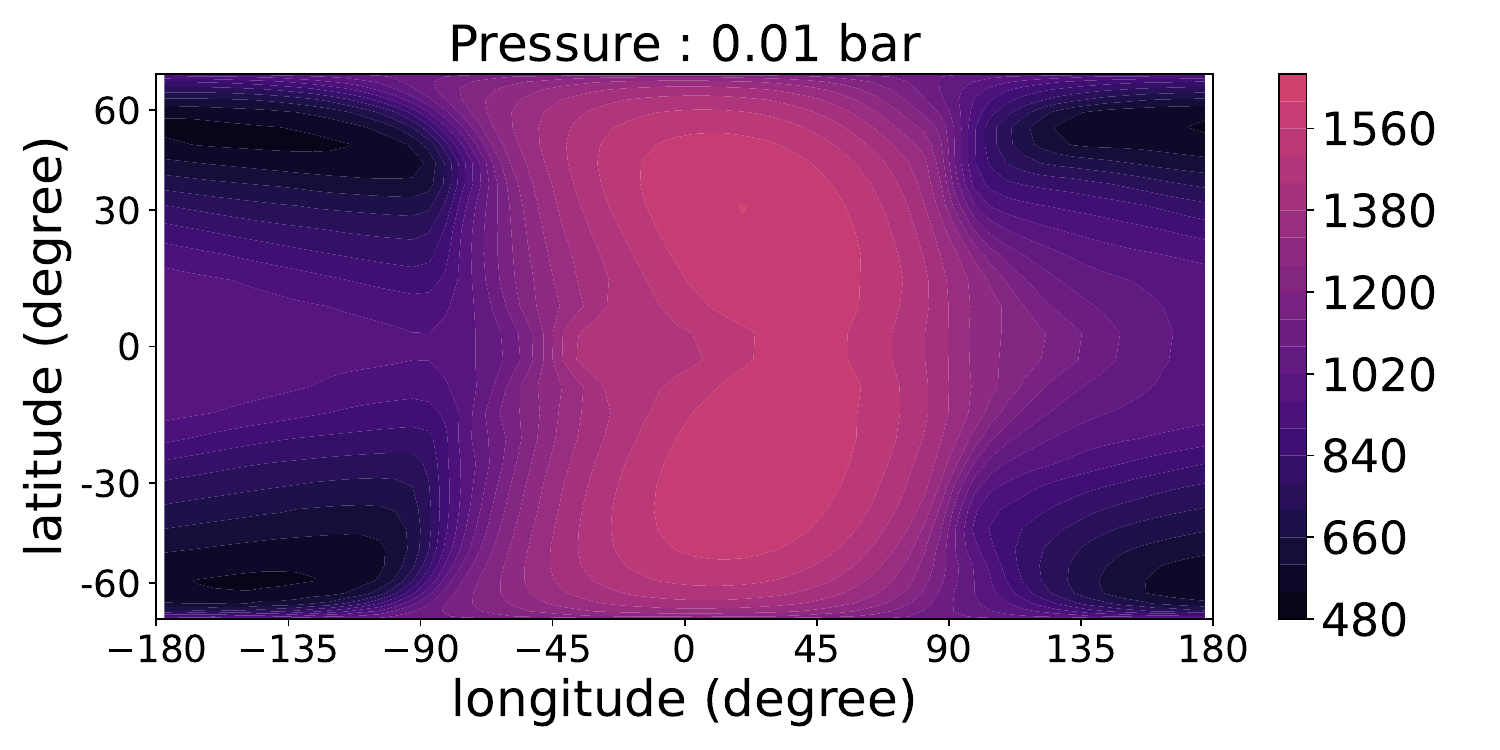}
\caption{Temperature (Kelvin) as a function of longitude and latitude at three pressure levels in the WASP-43b GCM. The super-rotating equatorial jet is clearly visible and shifts the `hot spot' eastward of the substellar point (where the star would be perceived to be directly overhead). Note that the substellar point is at 0 degree longitude. Such jet-like features would cause the phase curve amplitudes to peak before secondary eclipses. Note that the latitudinal distance is weighted by cos(latitude) to mimic the effect that polar latitudes would appear foreshortened to us because we observe WASP-43b from above the equator. }
\label{fig:gcm}
\end{figure}

\begin{figure}
\includegraphics[width=\columnwidth]{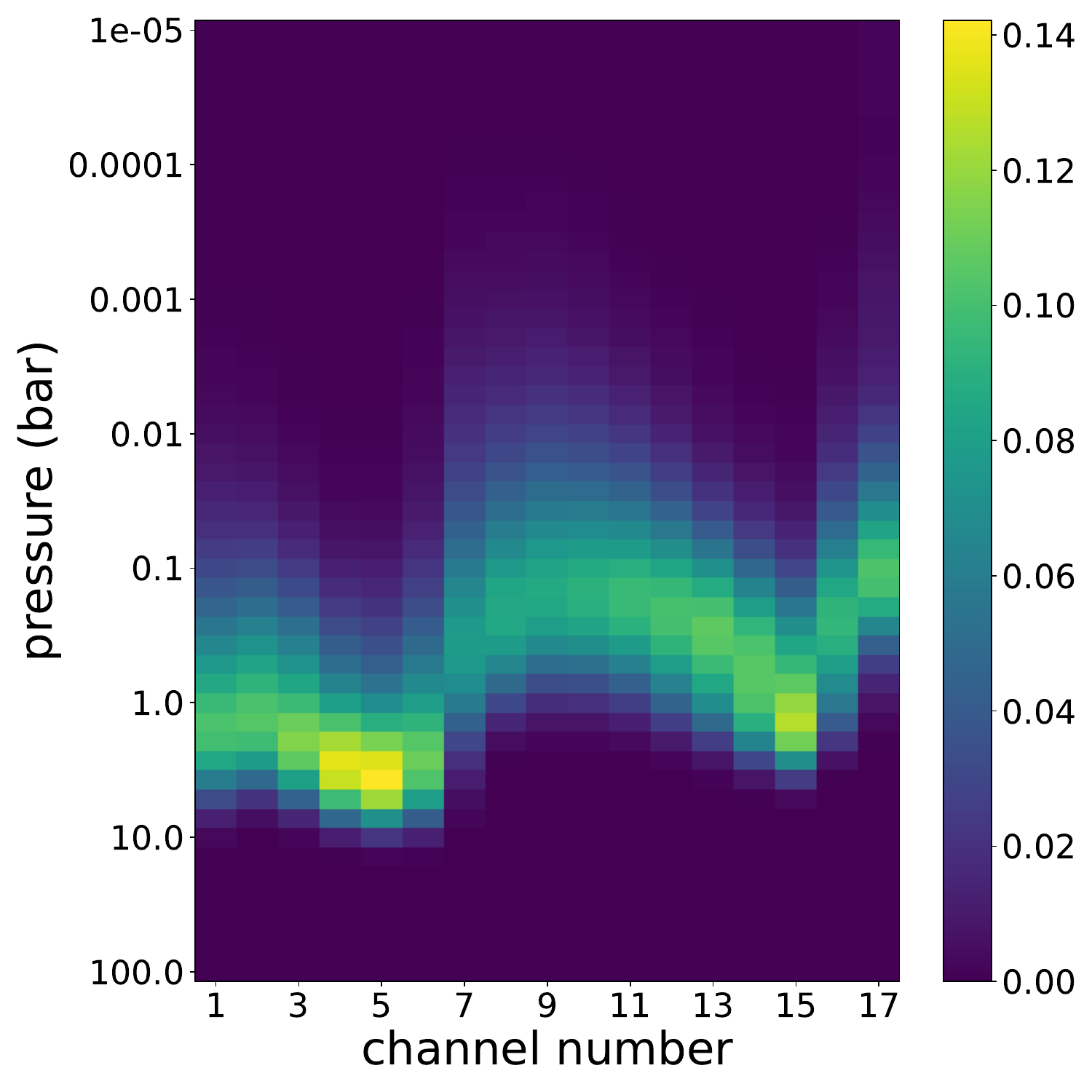}
\caption{Transmission weighting function of the WASP-43b GCM atmosphere at the substellar point as a function of pressure and wavelength channel. The first 15 \textit{HST}/WFC3 channels are in equally spaced bins of width 0.035 $\upmu$m spanning the wavelength range 1.1425-1.6325 $\upmu$m, and the last two \textit{Spitzer}/IRAC broad channels are centred at 3.6 and 4.5 $\upmu$m, respectively.}
\label{fig:cwf}
\end{figure}

\begin{table}
\centering
\caption{Gas volume mixing ratios (VMRs) and opacity data used to simulate synthetic phase curves. The VMRs are uniform with longitude, latitude and altitude. For He and H$_2$ collision-induced absorption opacity, we use the  coefficients from \protect\cite{borysow_collision-induced_1989} and \protect\cite{borysow_collision-induced_1989-1}.}
\label{tab:gcm_vmrs}
\begin{tabular}{lccc}
    \hline
    Molecule & VMR &  Opacity Data  \\
    \hline
    H$_2$O & $4.8\times10^{-4}$ & \cite{barber_high-accuracy_2006}    \\
    CO$_2$ & $7.4\times10^{-8}$ & \cite{tashkun_cdsd-4000_2011}    \\
    CO     & $4.6\times10^{-4}$ & \cite{rothman_hitemp_2010}    \\
    CH$_4$ & $1.3\times10^{-7}$ & \cite{yurchenko_exomol_2014}   \\
    He     & 0.162              &     \\
    H$_2$  & 0.837              &     \\
    \hline
\end{tabular}
\end{table}
We use a GCM of WASP-43b to simulate synthetic phase curves to validate our retrieval schemes.
We can directly assess the performance of our retrieval schemes by comparing the atmospheric properties retrieved from the synthetic data with the input GCM. 
The GCM is a cloud-free model calculated using SPARC/MITgcm \citep{showman_atmospheric_2009} based on the set-up of \cite{parmentier_transitions_2016}, and used for validating the 2.5D retrieval scheme by \cite{irwin_25d_2020}. 
We plot temperature as a function of longitude and latitude at three pressure levels of the GCM in Fig. \ref{fig:gcm}, and plot the transmission weighting function at the substellar point as a function of pressure and wavelength channel number in Fig. \ref{fig:cwf}.
The model is H$_2$/He dominated and contains four spectrally active gas species: H$_2$O, CO, CO$_2$ and CH$_4$, which are expected to be the dominant opacity sources in the atmosphere of WASP-43b in the observed wavelengths. 
The chemical abundance in the GCM is initially set according to chemical equilibrium, resulting in significant variation in CH$_4$ abundance from dayside to nightside in the photospheric pressures.
However, disequilibrium chemistry processes such as horizontal quenching are expected to smooth out such inhomogeneity \citep{cooper_dynamics_2006, agundez_pseudo_2014}.
Hence, we reset the abundance of all molecules to be the latitudinally averaged abundances in the 0.1 to 1-bar pressure region (using cos(latitude) as the weight) at the sub-stellar meridian, following \cite{irwin_25d_2020}. 
We then use this model to simulate the synthetic data. 
By resetting the chemical abundance to uniform values, we isolate the effect of temperature parameterisation in retrievals. 
We discuss the motivation and implication of using uniform abundance in section \ref{sec:lat_mean_vmr} and section \ref{limitation_chemistry}.
The volume mixing ratios (VMRs) in our GCM-based model are presented in Table \ref{tab:gcm_vmrs}. 
The synthetic phase curves are thus generated from the thermal structure of the GCM and the uniform VMRs, using the disc-averaging scheme described in section \ref{sec:disc-average}. 
The chemical abundance and thermal structure of this GCM-based model are seen as the `ground-truths' for our retrieval tests in section \ref{sec:gcm_test}.

\subsection{Radiative transfer calculation}
\label{sec:radiative_transfer}

We use the correlated-k method \citep{lacis_description_1991} to accurately and efficiently implement our radiative transfer calculations, following \cite{irwin_nemesis_2008}. 
Consider the mean transmission for a homogeneous path of absorber amount $m$ in the wavelength bin $[\lambda,\lambda+\Delta \lambda]$, 
\begin{equation}
    \overline{T}(m) = \f{1}{\Delta \lambda } \int^{\lambda + \Delta \lambda } _{\lambda} \exp(-k(\lambda) m) d\lambda,
    \label{eq:mean_transmission}
\end{equation}
where $k(\lambda)$ is the absorption cross-section. $\overline{T}$ is the key quantity that links opacity structure and thermal structure in thermal emission calculations. The cross-section $k$ is a rapidly varying function of $\lambda$, so it is computationally expensive to numerically calculate equation (\ref{eq:mean_transmission}). 
However, since the ordering of $k$ in the wavelength bin $[\lambda,\lambda+\Delta \lambda]$ does not affect the value of equation (\ref{eq:mean_transmission}), we sort $k$ in ascending order within each wavelength bin, which gives a monotonic distribution of $k$ that is easier to handle in quadrature schemes. 
Mathematically, let the cumulative frequency distribution of $k$ be $g(k)$, then the inverse of $g(k)$, which we denote as $k(g)$, is well-defined and monotonic. 
The function $k(g)$ is called the $k$-distribution, and can be tabulated on a grid of pressures and temperatures for each spectrally active molecule before calculations. 
During radiative transfer calculations, the $k$-distributions of multiple gases are combined with the random-overlapping-line approximation \citep{lacis_description_1991}. 
Such approximation gives residuals insignificant compared with measurement error, as found by \cite{irwin_25d_2020} and \cite{molliere_model_2015}. 
To calculate $\overline{T}$ through an inhomogenous path, we first split the path into multiple sub-paths \citep{irwin_nemesis_2008} that are sufficiently homogeneous, then model each sub-path with the absorber-amount weighted averaged sub-path properties. In the monochromatic case, the transmission of each sub-path can be multiplied together to give the transmission of the total path. 
However, to use the $k$-distribution technique where we have reordered $k$, we need to additionally assume that the wavelength regions of high opacity are correlated for all sub-paths, which is the correlated-k approximation. This is a good approximation for our set-up, as we are assuming constant vertical distribution of chemical abundance in our atmospheric model. 

Using the $k$-distribution technique, the mean transmission of a sub-path as defined in equation (\ref{eq:mean_transmission}) can be well-approximated with a Gaussian quadrature scheme:
\begin{equation}
    \overline{T}(m) = \sum_{i=1}^{N_g} \exp(-k_im)\Delta g_i,
    \label{eq:mean_transmission_quadrature}
\end{equation}
where $k_i$ is the $i$th quadrature point, $\Delta g_i$ the corresponding weight, and $N_g$ the number of quadrature points. The total transmission of an inhomogenous path is then 
\begin{equation}
    \overline{T}(m) = \sum^{N_g}_{i=1}\exp\big(-\sum^{N_{\tx{layer}}}_{j=1} k_{ij}m_j\big)\Delta g_i,
    \label{eq:total_transmission_quadrature}
\end{equation}
where we have multiplied the transmission of all $N_{\tx{layer}}$ layers together.
We use $k$-distribution look-up tables (`$k$-tables') with $N_g=20$ generated from the line data summarised in Table \ref{tab:gcm_vmrs}.

To speed up calculations, we use channel-averaged $k$-tables for the 15 channels of \textit{HST}/WFC3 and the 2 channels of \textit{Spitzer}/IRAC. \cite{irwin_25d_2020} find that such an approach produces an excellent approximation, resulting in residuals much less than typical measurement uncertainties of observations using these facilities. 
Apart from the above molecular opacity, we additionally include collision-induced absorption of H$_2$-H$_2$ pairs and H$_2$-He pairs using the coefficients of \cite{borysow_collision-induced_1989} and \cite{borysow_collision-induced_1989-1}, as well as Rayleigh scattering for a H$_2$/He dominated atmosphere using data from \cite{allen_astrophysical_1976}.
 
\subsection{Disc-average scheme}
\label{sec:disc-average}
\begin{figure}
\includegraphics[width=\columnwidth]{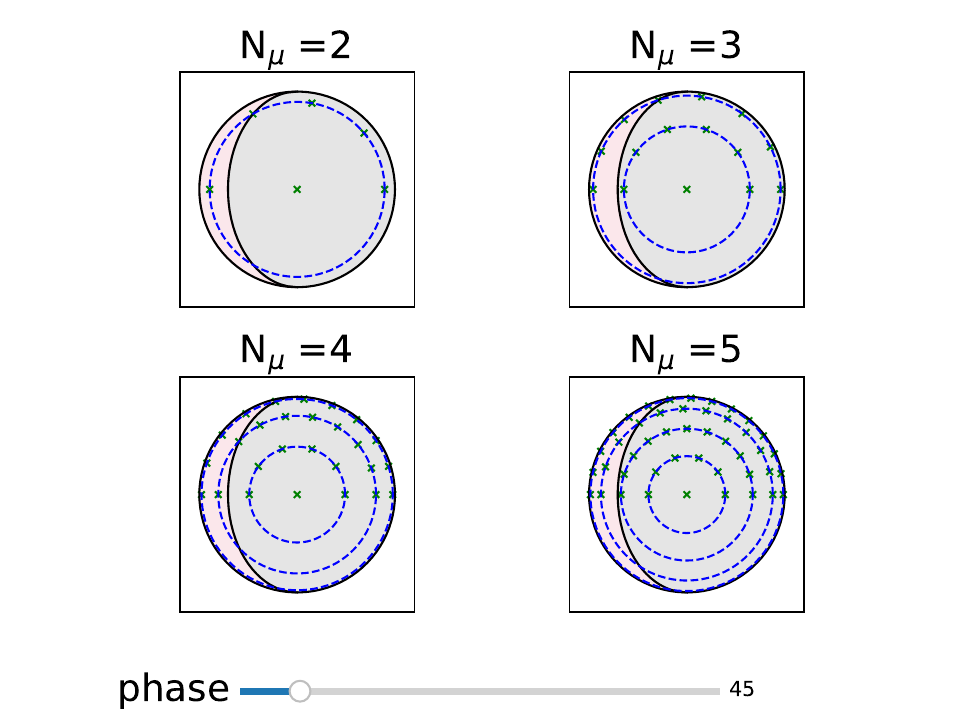}
\caption{Illustration of our disc averaging scheme at 45 degree orbital phase (0 degree being the primary transit and 180 degree the secondary eclipse). The visible region of the illuminated dayside at this orbital phase is shaded pink, whereas the visible nightside region is shaded in grey. The crosses mark the quadrature points for disc integration, and the dashed circles mark the positions of the zenith angle quadratures.}
\label{fig:disc_average_scheme}
\end{figure}

The disc-averaged spectral radiance (W m$^{-2}$ sr$^{-1}$ $\upmu$m$^{-1}$) of an inhomogeneous atmosphere for a distant observer is
\begin{equation}
\bar{R} (\lambda) =  \f{1}{\pi} \int^{2\pi}_{\phi=0}\int^{1}_{\mu=0} R(\lambda, \mu, \phi)\mu d\mu d\phi , 
    \label{eq:disc_averaged_radiance}
\end{equation}
where $\mu = \cos(\theta$) is the cosine of the zenith angle\footnote{The zenith angle is defined as the angle between the local normal of the atmosphere and the line of sight.} $\theta$, and $\phi$ is the azimuth angle. 
To carry out sampling-based Bayesian parameter estimation, we need to evaluate $\bar{R} (\lambda)$ for many different model atmospheres to approximate the posterior distributions, so it is important to have a numerical integration scheme for equation (\ref{eq:disc_averaged_radiance}) that is both accurate and computationally inexpensive. 
We use the method of \cite{irwin_25d_2020}: the zenith integration with respect to $\mu$ is done with a Gauss-Lobatto quadrature scheme with $N_{\mu}$ quadrature points, while the azimuthal integration with respect to $\phi$ is done with a Trapezium rule quadrature scheme with $N_{\phi}$ quadrature points. 
For the trapezium rule integration, the quadrature points are placed on the circles corresponding to each zenith quadrature point such that the arc-length between neighbouring points is $\sim$ $R_{\tx{plt}}/N_{\mu}$ (Fig. \ref{fig:disc_average_scheme}). 
Overall, our numerical integration scheme for equation (\ref{eq:disc_averaged_radiance}) is given by: 
\begin{equation}
\bar{R} (\lambda) = 2\sum^{N_{\mu}}_i \sum^{N_{\phi}}_j R(\lambda, \mu_i, \phi_{ij}) \mu_i \Delta \mu_i w_{ij},
\label{eq:numerically_averaged_radiance}
\end{equation}
where $\mu_i$ are the Lobatto quadrature points for the zenith integration,  $\Delta \mu_i$ are the corresponding quadrature weights, $\phi_{ij}$ are the Trapezium rule quadrature points for the azimuth integration (which are different for each zenith angle), and $w_{ij}$ are the Trapezium rule quadrature weights for the $j$th azimuthal angle and the $i$th zenith angle. 
We assume that the orbit of the planet is exactly edge-on, and that the atmosphere is symmetric about the equatorial plane, which is defined to be in the orbital plane. As a result, we only need to evaluate the integration over half of the visible disc and multiply the result by a factor of two. The quadrature points for $N_{\mu}=2,3,4,5$ are shown in Fig. \ref{fig:disc_average_scheme}.

For atmospheric models that partition the atmosphere in longitude into several regions each modelled with a single TP and abundance profile, our disc average routine can be further simplified. 
For example, if the planet is divided into a uniform dayside and a uniform nightside, as illustrated in Fig. \ref{fig:disc_average_scheme}, then all the quadrature points on the same zenith angle ring (blue dashed circles) in the same region have the same radiance. 
The azimuth integration is then a matter of calculating what fraction of the zenith angle rings are in each region. 
This greatly speeds up the disc averaging routine for the simple models in \ref{sec:simple_T_models}.

\subsection{2D atmospheric temperature models}
\label{sec:simple_T_models}
        
\begin{figure}
\includegraphics[width=\columnwidth]{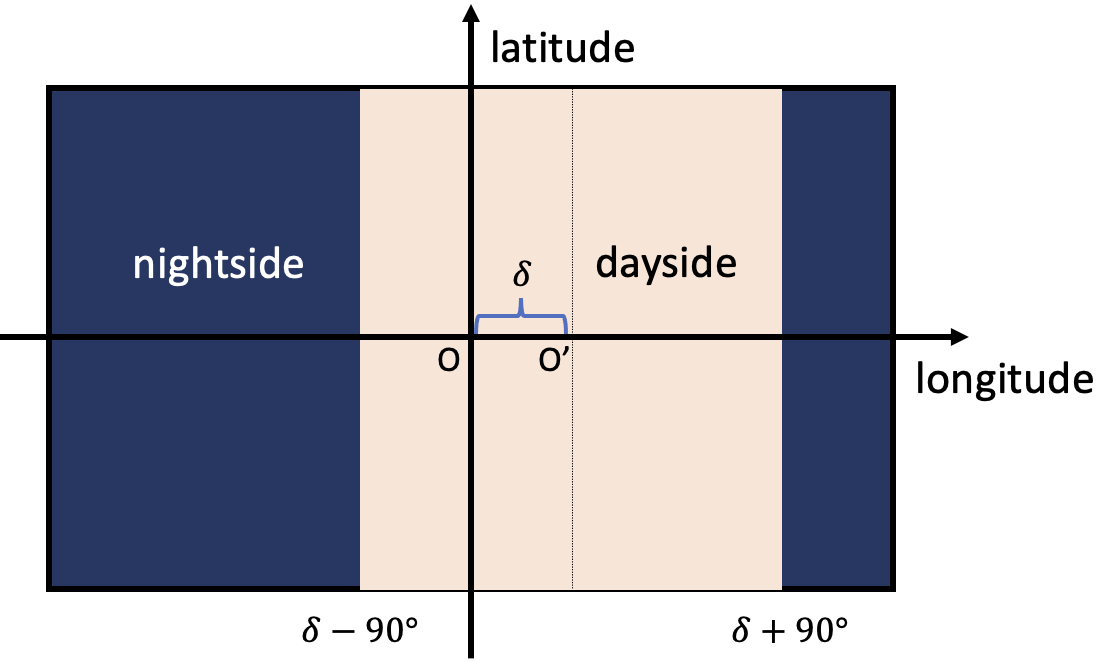}
\includegraphics[width=\columnwidth]{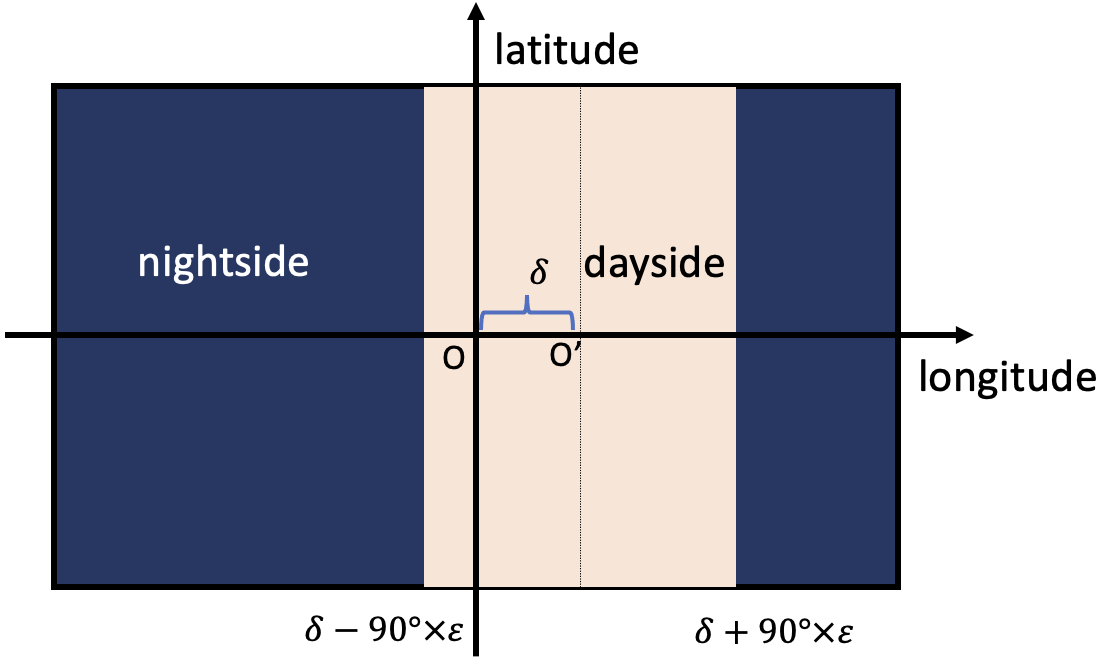}
\caption{Schematics of model 1 and model 2. Model 1 (top panel) is defined by equation (\ref{eq:model1}) and divides the atmosphere into a dayside and a nightside. Each region is then modelled with a single representative TP profile. The dayside central longitude $\delta$ is allowed to vary, and the dayside width (longitudinal extent) is fixed to be 180$^{\circ}$. Model 2 (bottom panel) is defined by equation (\ref{eq:model2}) and generalises model 1 by allowing the dayside width to vary, which now spans 180$^{\circ}\times\varepsilon$ in longitude.}
\label{fig:model_1_2}
\end{figure}

\begin{figure}
\includegraphics[width=\columnwidth]{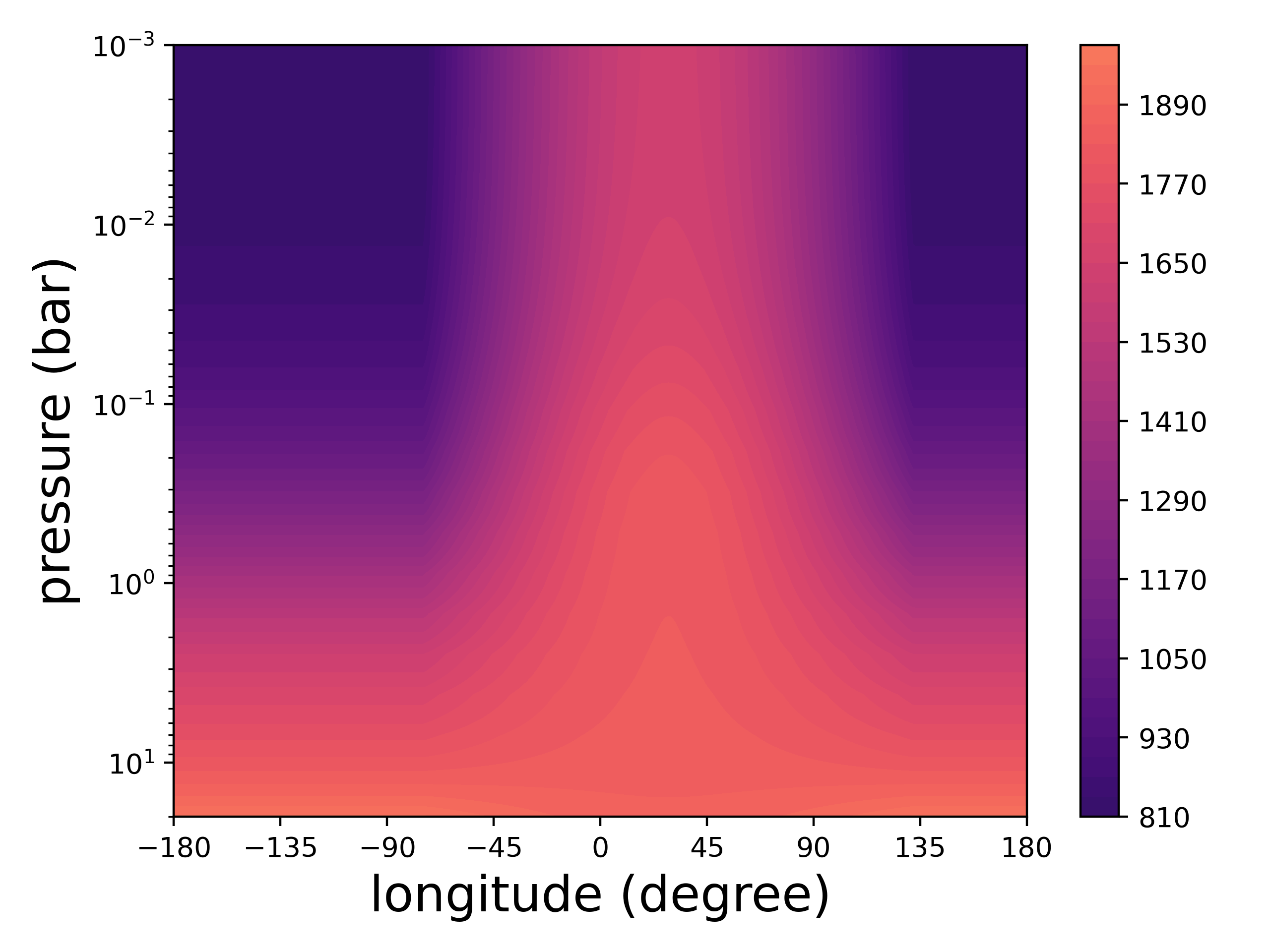}
\includegraphics[width=\columnwidth]{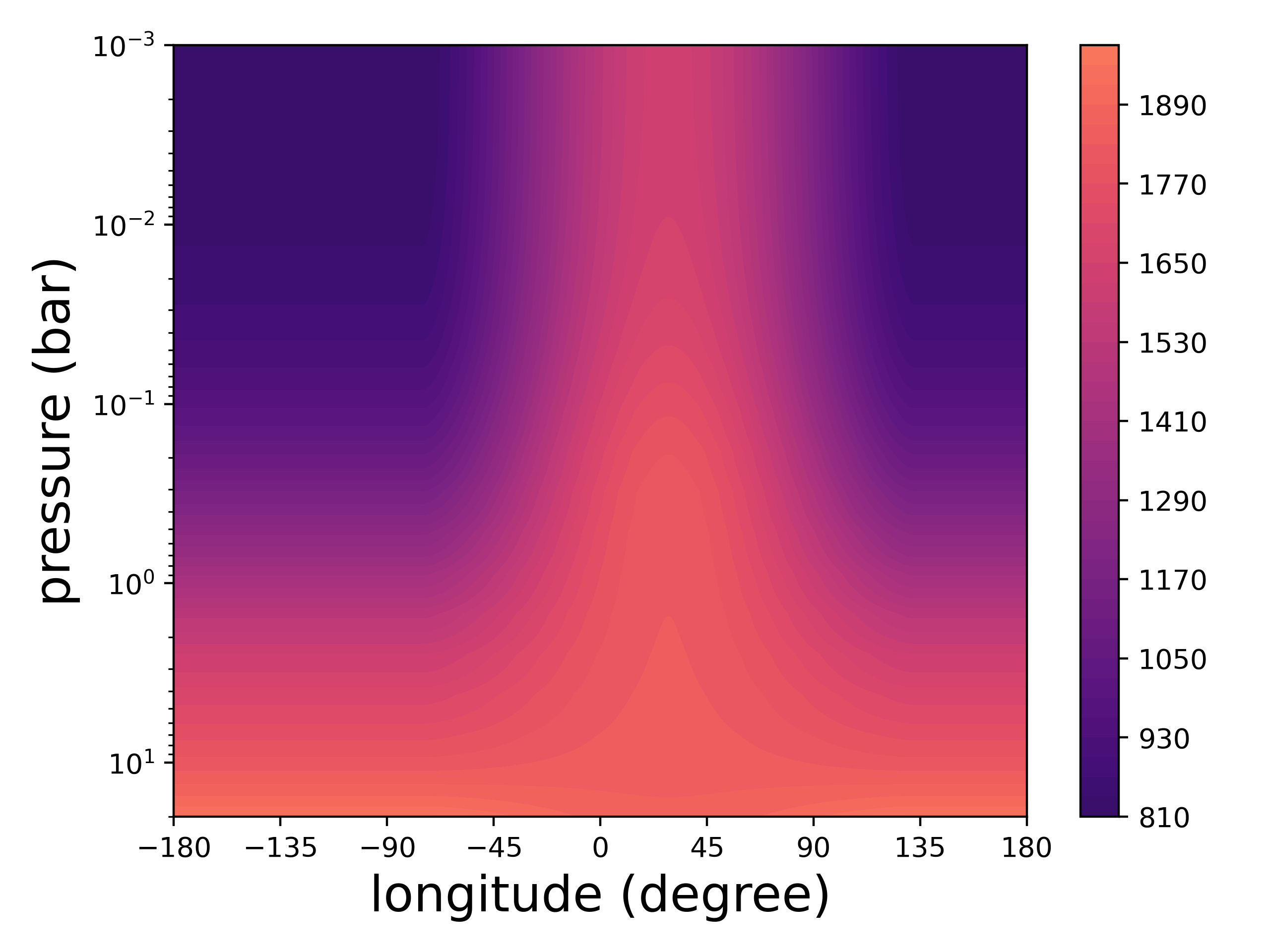}
\caption{Temperature as a function of pressure and longitude in two examples of model 3 and model 4. Model 3 (top panel) is defined by equation (\ref{eq:model3}) . 
Model 4 (bottom panel) is defined by equation (\ref{eq:model4}); in this example $n$ is set to be 1.75. Note that model 4 is equivalent to model 3 if $n$ is being set to 1. }
\label{fig:model_3_4}
\end{figure}
        
We describe four parametric temperature models for the atmospheres of hot Jupiters. 
Model 4 is our proposed model, and the other three simpler models are included for comparison.
While all of the models use the one-dimensional analytical TP profile of \cite{guillot_radiative_2010} to describe temperature as a function of pressure, they can also be easily interfaced with other parametric TP profiles. 
The Guillot profile is given by equation (29) of \cite{guillot_radiative_2010}
\begin{equation}\label{2_stream}
T^4 = \f{3 T_{\tx{int}}^4}{4} \Big( \f{2}{3} + \tau \Big) +  \f{3 T_{\tx{irr}}^4}{4} f \Big[ \f{2}{3} + \f{1}{\gamma\sqrt{3}} + \Big( \f{\gamma}{\sqrt{3}} - \f{1}{\gamma \sqrt{3}} \big) \tx{e}^{-\gamma \tau \sqrt{3}}\Big],
\end{equation}
where $\tau$ is the infrared optical depth defined by
\begin{equation}
    \tau(P) = \f{\kappa_{\tx{th}} P}{g}.
\end{equation}
The TP profile has four free parameters: $\kappa_{\tx{th}}$ is the mean infrared opacity, $\gamma$ is the ratio between the mean visible and mean infrared opacities, T$_{\tx{int}}$ is the internal heat flux, and $f$ is a catch-all parameter of order unity that models the effects of albedo (on the dayside) and the redistribution of stellar flux due to atmospheric circulation (on both the dayside and the nightside). 
We assume the change in gravity $g$ is negligible in the pressure range probed by emission spectroscopy, so that  $\tau$ is linear in $P$. Finally, $T_{\tx{irr}}$ is the irradiation temperature defined by
\begin{equation}
T_{\tx{irr}} = \Big(\f{R_{\tx{star}}}{a}\Big)^{1/2} T_{\tx{star}},
\end{equation}
where $a$ is the orbital semi-major axis, and $R_{\tx{star}}$ and $T_{\tx{star}}$ are the host star radius and temperature.
In section \ref{sec:1Dfit}, we show that the Guillot profile is able to approximate the temperature-pressure profiles found in the WASP-43b GCM well enough to reproduce the synthetic data.
For all models, we place the sub-stellar point at the origin in our longitude-latitude coordinate system. 
We denote longitude by $\Lambda$. 
All models describe the thermal structure with two TP profiles: a representative dayside profile and a representative nightside profile.
Note that temperature is set to be uniform as a function of latitude on isobars, and we treat the retrieved temperature profiles as latitudinally averaged temperature profiles. 
In our models, the centres of `dayside' and the `nightside' can shift away from the substellar point and the anti-stellar point, respectively.

\subsubsection{Model 1}
Model 1 (Fig. \ref{fig:model_1_2}, top panel) our simplest model in which the dayside and nightside are both set to span 180 $^{\circ}$ in longitude.  
The centre of the dayside region ($O'$) is allowed to shift eastward or westward by some longitude $\delta$ relative to the substellar point, representing the effect of atmospheric dynamics, in particular equatorial jets, on redistributing heat around the atmosphere. 
The dayside is thus bound by the meridians $\Lambda=\delta-90^{\circ}$ and  $\Lambda=\delta+90^{\circ}$. 
Within each region, temperature is only a function of pressure. 
Note that in all our models, the `dayside' denotes the region of the atmosphere modelled by the dayside TP profile and does not necessarily coincide with the physically illuminated dayside. 
The model contains 9 parameters: 4 parameters for each of the two TP profiles and 1 parameter $\delta$ for the longitudinal offset. 
Note that model 1 is equivalent to the `2TP-Crescent' approach of \cite{feng_2d_2020} when applied to all phases simultaneously. 
In summary, model 1 is given by:
\begin{equation}
\label{eq:model1}
T =
\begin{cases}
    T_{\tx{night}}(P) & \text{if }\Lambda > \delta + 90^{\circ} \text{ or } \Lambda < \delta - 90^{\circ},\\
    T_{\tx{day}}(P) & \text{otherwise.}
\end{cases}
\end{equation}

\subsubsection{Model 2}
Model 2 (Fig. \ref{fig:model_1_2}, lower panel) is similar to model 1, with the only difference being that the `width' (longitudinal extent) of the dayside is now a free parameter. 
We introduce a scaling parameter $\varepsilon$, so that the dayside is now bounded by the meridians $\Lambda=\delta-90^{\circ}\times\varepsilon$ and $\Lambda=\delta+90^{\circ}\times\varepsilon$ and spans 180$^{\circ}\times\varepsilon$ in longitude. The model contains 10 parameters: 4 parameters for each TP profile, 1 parameter $\delta$ for the longitudinal offset, and 1 parameter $\varepsilon$ for the dayside width.
Parameterising the dayside area fraction has been shown to be effective in analysing disc-averaged emission spectrum of tidally-locked hot Jupiters by \cite{taylor_understanding_2020}, and this approach has been applied to phase curve analysis by \cite{feng_2d_2020} in their `2TP-Free' model, albeit only in the phase-by-phase approach. Model 2 is our way of implementing the dayside area fraction parameterisation self-consistently when fitting all phases of phase curves simultaneously. In summary, model 2 is given by:
\begin{equation}
\label{eq:model2}
T =
\begin{cases}
    T_{\tx{night}}(P) & \text{if } \Lambda >\delta + 90^{\circ}\varepsilon \text{ or } \Lambda <\delta - 90^{\circ}\varepsilon,\\
    T_{\tx{day}}(P) & \text{otherwise.}
\end{cases}
\end{equation}

\subsubsection{Model 3}
Model 3 (Fig. \ref{fig:model_3_4}, upper panel) is an extension of model 2, where the temperature is now a continuous function of longitude across the dayside boundary.
The dayside is bound by the meridians $\Lambda=\delta-90^{\circ}\times\varepsilon$ and $\Lambda=\delta+90^{\circ}\times\varepsilon$. Within the dayside, temperatures at each pressure level vary with the cosine of longitude. Outside the dayside, the TP is set to be a single nightside profile. In summary, model 3 is given by:
\begin{equation}
\label{eq:model3}
T =  
\begin{cases} 
T_{\tx{night}}(P) \text{ if }\Lambda > \delta + 90^{\circ} \text{ or } \Lambda < \delta - 90^{\circ}, \\
T_{\tx{night}}(P) + (T_{\tx{day}}(P)-T_{\tx{night}}(P))\cos(\f{\Lambda-\delta}{\varepsilon})  \text{ otherwise.}
\end{cases}
\end{equation}

\subsubsection{Model 4}
Model 4 is a generalisation of model 3, where we allow the exponent of the cosine term in equation (\ref{eq:model3}) to be a variable, so that we parameterise how strongly temperatures vary with longitude on isobars:
\begin{equation}
\label{eq:model4}
T =  
\begin{cases} 
T_{\tx{night}}(P) \text{ if }\Lambda > \delta + 90^{\circ} \text{ or } \Lambda < \delta - 90^{\circ}, \\
T_{\tx{night}}(P) + (T_{\tx{day}}(P)-T_{\tx{night}}(P))\cos^n(\f{\Lambda-\delta}{\varepsilon})  \text{ otherwise.}
\end{cases}
\end{equation}
The parameters for the temperature models, together with the other parameters of our atmospheric models, are summarised in Table \ref{tab:priors}.

\subsection{Retrieval set-up}
\label{sec:retrieval_set_up}

\begin{table*}
\centering
\caption{Parameters of our atmospheric models. All models have 4 parameters for gas VMRs, with the other parameters specifying the thermal structure. Model 1 has 13 parameters in total, whereas model 2 and model 3 have 14 parameters in total. Model 4 has 15 parameters in total.}
\label{tab:priors}
\begin{tabular}{lccc}
    \hline
    Parameter & Description & Model Usage & Range \\
    \hline
    $\delta$ & Dayside longitudinal offset  & all  & [-45$^{\circ}$,45$^{\circ}$]        \\
    $\varepsilon$ & Dayside longitudinal width scaling & 2,3,4                 & [0.5,1.2]   \\
    $n$ & Dayside longitudinal variation exponent & 4 & [0,2] \\
    $\log \kappa_{\tx{th,day}}$ & Mean infrared opacity (dayside)  & all  & [-4,2]   \\
    $\log\gamma_{\tx{day}}$ &  Ratio of visible and infrared opacities (dayside) & all  &  [-4,1]      \\
    $\log f_{\tx{day}}$ & Heat redistribution parameter (dayside) & all    & [-4,1]  \\
    T$_{\tx{int,day}}$ &  Internal heat flux (dayside)   & all  &   [100,1000]               \\
    $\log \kappa_{\tx{th,nigtht}}$  & Mean infrared opacity (nightside)  & all    & [-4,2]      \\
    $\log\gamma_{\tx{night}}$ & Ratio of visible and infrared opacities (nightside) & all     & [-4,1]     \\
    $\log f_{\tx{night}}$ & Heat redistribution parameter (nightside) & all    &   [-4,1]    \\
    T$_{\tx{int,night}}$ &  Internal heat flux (nightside)  & all    &  [100,1000]  \\
    log VMR$_{\tx{H$_2$O}}$ & Log$_{10}$ volume mixing ratio of H$_2$O & all & [-8,-2]    \\
    log VMR$_{\tx{CO$_2$}}$ & Log$_{10}$ volume mixing ratio of CO$_2$ & all & [-8,-2]    \\
    log VMR$_{\tx{CO}}$ & Log$_{10}$ volume mixing ratio of CO & all & [-8,-2]    \\
    log VMR$_{\tx{CH$_4$}}$ & Log$_{10}$ volume mixing ratio of CH$_4$ & all  & [-8,-2]    \\
    \hline
\end{tabular}
\end{table*}
We run retrievals on two sets of data: (1) synthetic \textit{HST}/WFC3 and \textit{Spitzer}/IRAC phase curves simulated from the GCM-based model of WASP-43b described in section \ref{sec:gcm_intro}, and (2) observed \textit{HST}/WFC3 and \textit{Spitzer}/IRAC phase curves of WASP-43b as presented in \cite{stevenson_spitzer_2017}. 
For each set of data, we run four retrievals using each of the atmospheric models described in section \ref{sec:simple_T_models}. 
We fit the spectra at all phases simultaneously using spectra generated from the parametric atmospheric models. 
We use Nested Sampling \citep{feroz_multimodal_2008} to calculate the posterior distribution of the atmospheric model parameters and the Bayesian evidence of the model, described in section \ref{sec:nested_sampling}.
In appendix \ref{app:phase_by_phase}, we also compare our retrieval results to the phase-by-phase retrieval approach, where the spectrum at each orbital phase is analysed independently.

For all of our retrievals, the atmospheric model is defined from 20 to $10^{-3}$ bar, on 20 points equally spaced in log pressure. The atmospheric models have two components: a temperature model and a chemical abundance model. In each of our retrieval schemes, we test a different temperature model described in section \ref{sec:simple_T_models}. 
On the other hand, all of our retrieval schemes share the same chemical abundance model, which assumes a H$_2$/He dominated atmosphere and contains four spectrally active gases: H$_2$O, CO$_2$, CO, CH$_4$. 
The abundance model is parameterised by the volume mixing ratios of the spectrally active gases, assumed to be constant with respect to pressure, longitude and latitude. 
Furthermore, we do not include clouds/hazes in any of our models, and we assume aerosols with no significant spectral features to be degenerate with the other components of our atmospheric models. 
The limitations of these assumptions are discussed in section \ref{sec:discussions}. 
The model parameters and their prior ranges are listed in Table \ref{tab:priors}, and we prescribe uniform priors for all of our model parameters. 
    
\subsection{Bayesian parameter estimation}
\label{sec:nested_sampling}
We extract information from phase curves using Bayesian inference. Consider a set of phase curve data $D$ that we wish to analyse. 
Suppose we have a parametric atmospheric model $M$ with parameter space $\Theta$, so that for each point $\theta\in\Theta$ we can calculate model phase curves $f(M(\theta))$, where $f$ is our `forward model' that encapsulates all the modelling steps required to generate model phase curves from an atmospheric model. 
The probability distribution of the parameters of $M$ given $D$ is  
\begin{equation}
    \Pr(\theta, M | D) =  \f{\Pr(D | \theta, M ) \Pr(\theta | M)}{\Pr(D)},
    \label{eq:bayes_theorem}
\end{equation}
where $P (\theta) \equiv \Pr(\theta, M | D)$ is the posterior distribution, $\Li(D) \equiv \Pr(D | \theta, M )$ is the likelihood,  $\pi(\theta)\equiv\Pr(\theta | M)$ is the prior, and $Z\equiv \Pr(D | M)$ is the evidence. To proceed, we define the log likelihood function as 
\begin{equation}
    \log \Li(\theta) = -\f{1}{2} \sum_{i=1}^{N_{\rm obs}} \f{(D_i - f(M(\theta)_i)^2}{\sigma_i^2}, 
    \label{eq:log_likelihood}
\end{equation}
where $N_{\rm obs}$ is the total number of data points in the observed phase curves, $D_i$ and $ f(M(\theta))_i$ are the $i$th points of the observed and model phase curves, respectively, and $\sigma_i$ is the associated measurement uncertainty. 
The Bayesian evidence is then given by 
\begin{equation}
    Z = \int_{\Theta} \Li(\theta) \pi(\theta) d^n\theta, 
    \label{eq:bayesian_evidence}
\end{equation}
where the integral is over the $n$-dimensional parameter space $\Theta$.
We can approximately calculate the posterior distribution and the evidence by sampling the parameter space $\Theta$, which is a computationally expensive task for phase curve retrievals because $\Theta$ is high-dimensional and $ f(M(\theta))$ is expensive to calculate. 
The Nested Sampling algorithm \citep{feroz_multimodal_2008} is an efficient way to carry out these tasks, which starts by rewriting the evidence in terms of the prior volume $X$, defined as 
    \begin{equation}
        X(\lambda) = \int_{\Li(\theta)>\lambda}\pi(\theta)d^n\theta.
    \end{equation}
The evidence is then given by 
    \begin{equation}
        \int_0^1 \Li(X)dX,  
    \end{equation}
where $\Li(X)$ is a monotonic function and can be evaluated with simple quadrature schemes, and the likelihood contours $\Li(X_i)$ are approximated by sampling the parameter space within nested ellipsoids. Numerically, the evidence is given by  
    \begin{equation}
        Z\approx\sum^{N_{\tx{iter}}}_{i=1}\Li(X_i)w_i,
    \end{equation}
where $X_i\in[0,1]$ are a decreasing sequence of quadrature points (starting from 1) and $w_i$ the corresponding weights, and $N_{\tx{iter}}$ is determined by some convergence criterion. The nested sampling algorithm discards the point with the lowest-likelihood at each iteration, so the posterior can be generated by assigning weights to those points by 
    \begin{equation}
        p_i = \f{\Li(X_i)w_i}{Z}.
    \end{equation}
For all of our retrievals, we use the Python interface PyMultiNest \citep{buchner_x-ray_2014} to implement nested sampling with 1000 sampling live points.

\section{Preliminary Tests}
\label{sec:preliminary}
Before testing our retrieval schemes on synthetic \textit{HST}/WFC3 and \textit{Spitzer}/IRAC phase curves, we investigate if simplified atmospheric models can reproduce the phase curves generated from the GCM described in section \ref{sec:gcm_intro}. 
We show that we can reproduce these GCM phase curves to well within realistic measurement uncertainties: (1) if we replace the TP profiles in the GCM at all locations with best-fit 1D Guillot profiles; (2) if we replace the TP profiles in the GCM on the same meridian with the latitudinally averaged TP profile of that meridian; and (3) if we replace the volume mixing ratios of all gases with a uniform profile. 
These results justify our use of 2D models coupled with the Guillot profile in our retrievals, and in section \ref{sec:gcm_test},  we show that such 2D models can adequately model synthetic \textit{HST}/WFC3 and \textit{Spitzer}/IRAC phase curves.

\begin{figure}
    \includegraphics[width=\columnwidth]{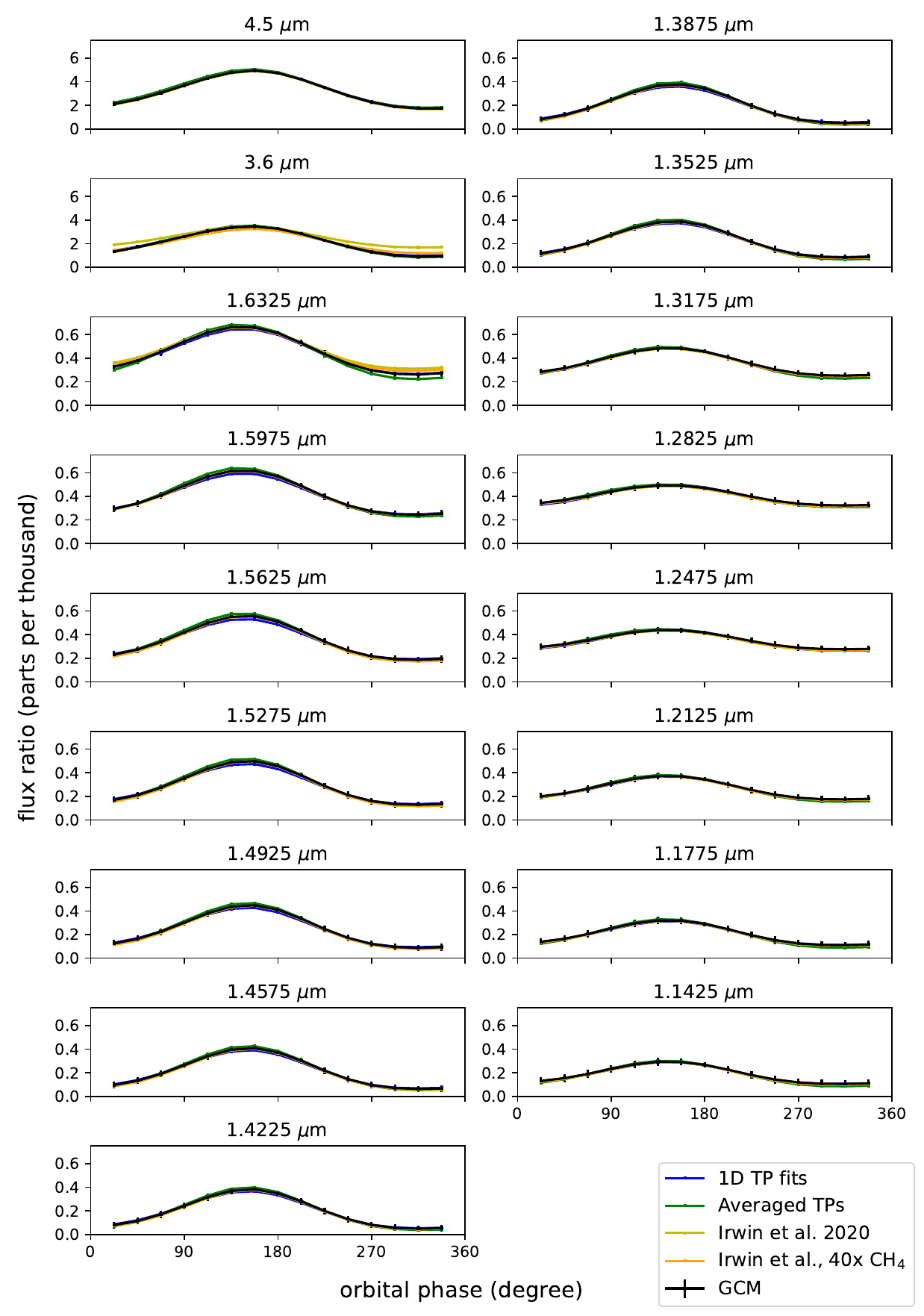}
    \caption{Comparison of the phase curves simulated from simplified models to the phase curves simulated from the original WASP-43b GCM (black), where abundance is set by chemical equilibrium. The blue curves are simulated from the best-fit Guillot profiles. The green curves are simulated from latitudinally-averaged TP profiles from 0$^{\circ}$ to 45$^{\circ}$, using cos(latitude) as the weight. The yellow curves are simulated with uniform abundance listed in Table \ref{tab:gcm_vmrs}, which are the synthetic data we retrieve on in section \ref{sec:gcm_test}.  The orange curves are simulated with the abundance listed in Table \ref{tab:gcm_vmrs}, but with the methane abundance multiplied by 40, which illustrates the fact that models with uniform gas abundance can match the phase curves produced from a chemical equilibrium model. Note that the error bars on the GCM phase curves are the estimated observational uncertainties of \protect\cite{stevenson_spitzer_2017}.}
    \label{fig:fit_1D_TP_to_gcm}
\end{figure}
 
\subsection{Replace GCM TP profiles with 1D model fits}
\label{sec:1Dfit}
We use the Guillot TP profile \citep{guillot_radiative_2010} as the basis of our 2D temperature models described in section \ref{sec:simple_T_models}. We demonstrate here that this profile is flexible enough to approximate the range of TP profiles found in our WASP-43b GCM with the following procedure. 
First, we directly fit the 1D TP profile to the TP profiles of the GCM on all longitude-latitude grid points in the pressure range 20-$10^{-3}$ bar, which covers the support of the transmission weighting function. 
We find that extending the pressure range has negligible effects on the spectra. 
We then generate phase curves from the total collection of best-fit 1D profiles, and compare them to those generated directly from the GCM. 
Both sets of phase curves are simulated using the volume mixing ratios of the original GCM, which are set via chemical equilibrium. 
In Fig. \ref{fig:fit_1D_TP_to_gcm}, we compare the phase curves simulated from the best-fit 1D profiles (blue curves) with the phase curves simulated directly from the GCM (black curves). 
We overplot the measurement errors of \cite{stevenson_spitzer_2017} on the phase curves simulated directly from the GCM. 
We see that the phase curves simulated from the 1D best-fit profiles can match the GCM phase curves to within error at almost all phases.

\subsection{Replace GCM thermal structure with latitudinally-averaged thermal structure}
\label{sec:lat_mean_tp}
    
\begin{figure}
    \includegraphics[width=\columnwidth]{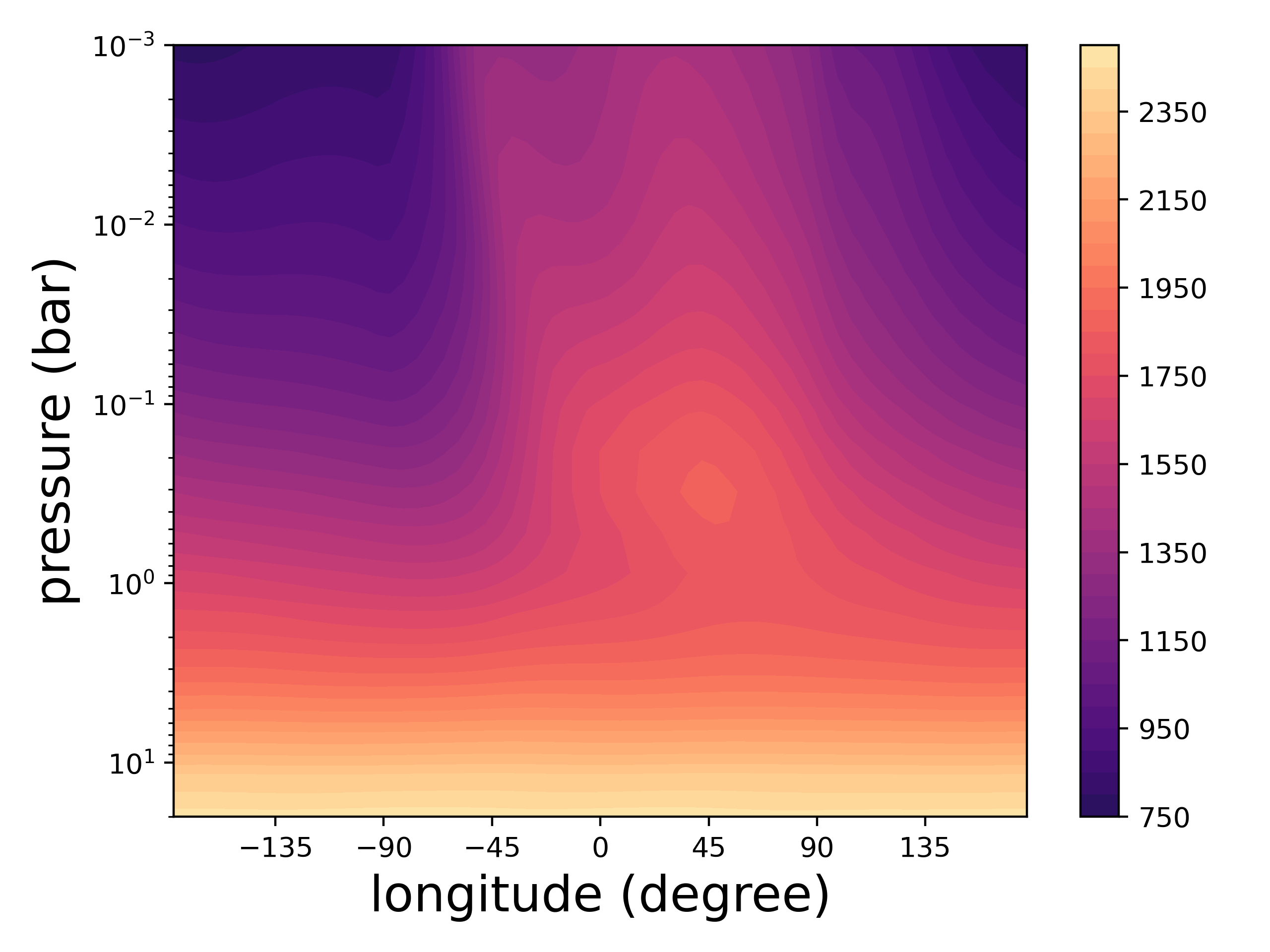}
    \includegraphics[width=\columnwidth]{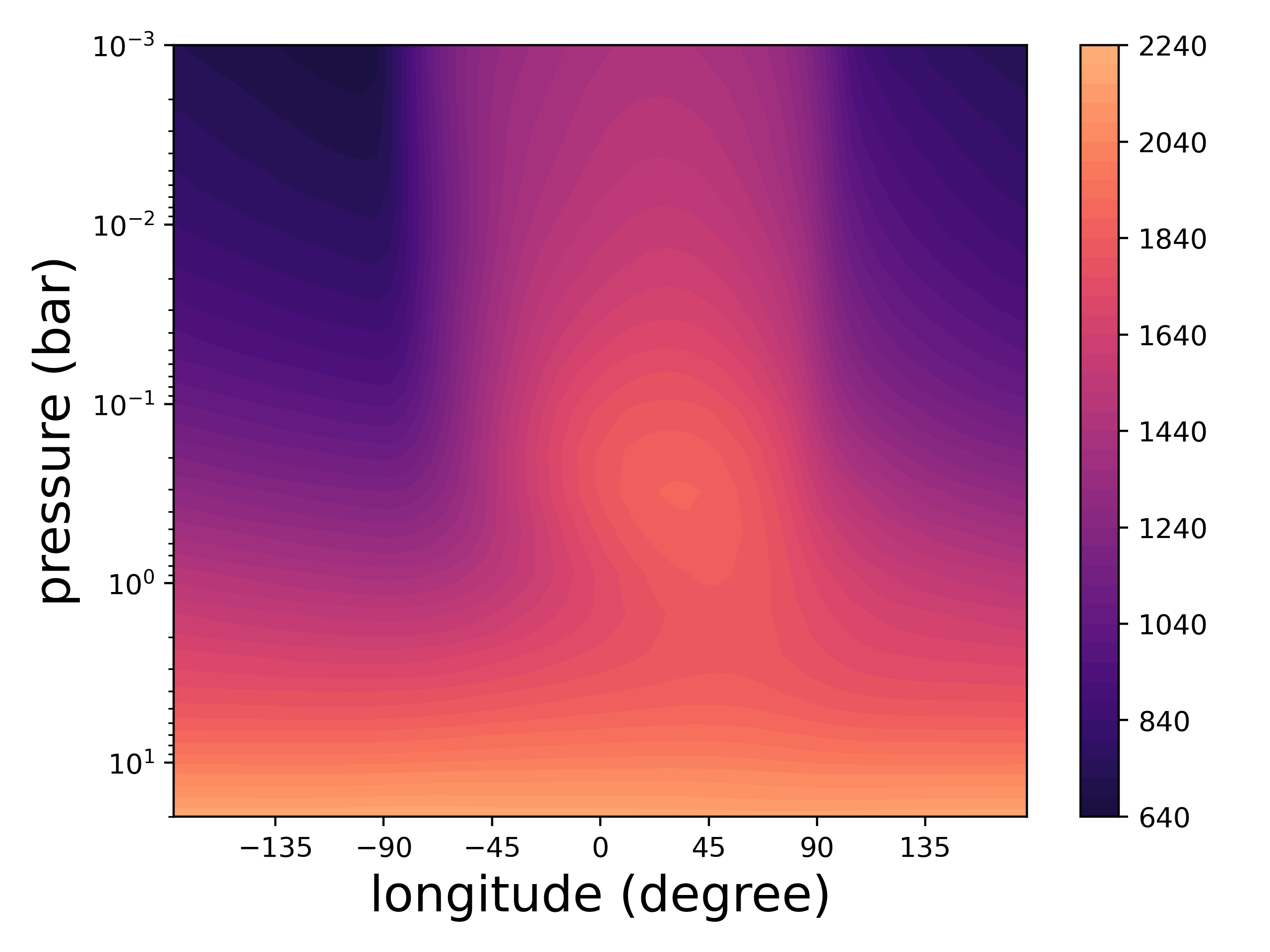}
    \caption{Top panel: temperature as a function of longitude and pressure in the WASP-43b GCM at the equator. Lower panel: latitudinally averaged TP profiles from 0 to 45 degree latitude, using cos(latitude) as the weight. Since the retrieved thermal structure resemble the latitudinally averaged profiles, we would underpredict the hot spot offset at the equator for typical hot Jupiter GCMs from synthetic data.}
    \label{fig:lat_mean_TP}
\end{figure}
In the 2D models described in section \ref{sec:simple_T_models}, the atmospheric temperature varies with longitude and pressure only,  and we prescribe that temperature is constant with respect to latitudinal variation. 
The key point is that we interpret the retrieved thermal structure as a latitudinally averaged thermal structure, as we have very limited sensitivity to latitudinal variation of atmospheric properties in the data.
We now demonstrate that the latitudinally-averaged thermal structure of the GCM can reproduce the synthetic data, which justifies our choice of 2D temperature models. 
To this end, we replace all the TP profiles on the same meridian with some latitudinally averaged TP structure. We find that if we pick the TP profile averaged from 0 degree latitude to 45 degree latitude using cos(latitude) as the weight, the resulting 2D temperature model could produce phase curves (green curves, Fig. \ref{fig:fit_1D_TP_to_gcm}) that agree with the phase curves simulated directly from the GCM (black curves, Fig. \ref{fig:fit_1D_TP_to_gcm}) to measurement uncertainties quoted in \cite{stevenson_spitzer_2017}. 
Note that both sets of phase curves are simulated using the chemical equilibrium VMRs of the original GCM. 
In conclusion, we find that a 2D temperature model can reproduce \textit{HST}/WFC3 and \textit{Spitzer}/IRAC quality phase curves simulated from our WASP-43b GCM. 
Furthermore, we expect the retrieved thermal structure using a 2D model to resemble the latitudinally averaged thermal structure of the atmosphere. 
In Fig. \ref{fig:lat_mean_TP}, we plot the GCM temperature as a function of pressure and longitude at the equator on the top, compared to the latitudinally averaged temperature (from 0 degree latitude to 45 degree latitude using cos(latitude) as the weight) on the bottom. 
In section \ref{sec:gcm_test}, we compare the retrieved thermal structure from the synthetic data with this latitudinally averaged GCM thermal structure.

\subsection{Replace chemical equilibrium VMRs with constant VMRs}
\label{sec:lat_mean_vmr}

As mentioned in section \ref{sec:gcm_intro}, the chemical equilibrium abundance of the original WASP-43b GCM is expected to be homogenised by horizontal quenching \citep{cooper_dynamics_2006, agundez_pseudo_2014}, and \cite{irwin_25d_2020} reset the GCM gas abundances at all altitudes and locations to be the latitudinally averaged abundances in the 0.1-1-bar pressure region (using cos(latitude) as the weight) at the sub-stellar meridian. 
We now compare the phase curves simulated using uniform abundance as those used by \cite{irwin_25d_2020} (yellow curves) with those simulated using the chemical equilibrium abundance (black curves) in Fig. \ref{fig:fit_1D_TP_to_gcm}.
We see the main difference is that the different distributions of CH$_4$ result in significantly different phase curves at 3.6 $\upmu$m. 
This effect has been investigated by \cite{steinrueck_effect_2019}, who explore if disequilibrium effects such as the quenching of CH$_4$ can explain why GCMs systematically overestimate phase curve amplitudes compared to observations. 
We echo the finding of \cite{steinrueck_effect_2019} that phase curves observed in the wavelength range covered by the 3.6 $\upmu$m \textit{Spitzer} channel are an effective diagnostic for disequilibrium methane chemistry on hot Jupiters.

The synthetic phase curves we use to validate our retrieval schemes are simulated from uniform gas VMRs as listed in Table \ref{tab:gcm_vmrs} and are the same synthetic phase curves in \cite{irwin_25d_2020}. 
To further justify our use of uniform gas abundance in our model, we show that if we multiply the CH$_4$ abundance of \cite{irwin_25d_2020} by a factor of 40, the resultant phase curves (orange curves) agree well with the phase curves simulated from the equilibrium chemistry VMRs (black curves). 
This suggests that while using uniform VMRs can adequately fit \textit{HST}/WFC3 and \textit{Spitzer}/IRAC quality phase curves, this approach can lead to significantly biased CH$_4$ abundance. 
The limitation of the constant chemistry assumption is discussed in \ref{limitation_chemistry}.

\section{Application to Synthetic Phase Curves}
    \label{sec:gcm_test}

\begin{figure}
    \includegraphics[width=\columnwidth]{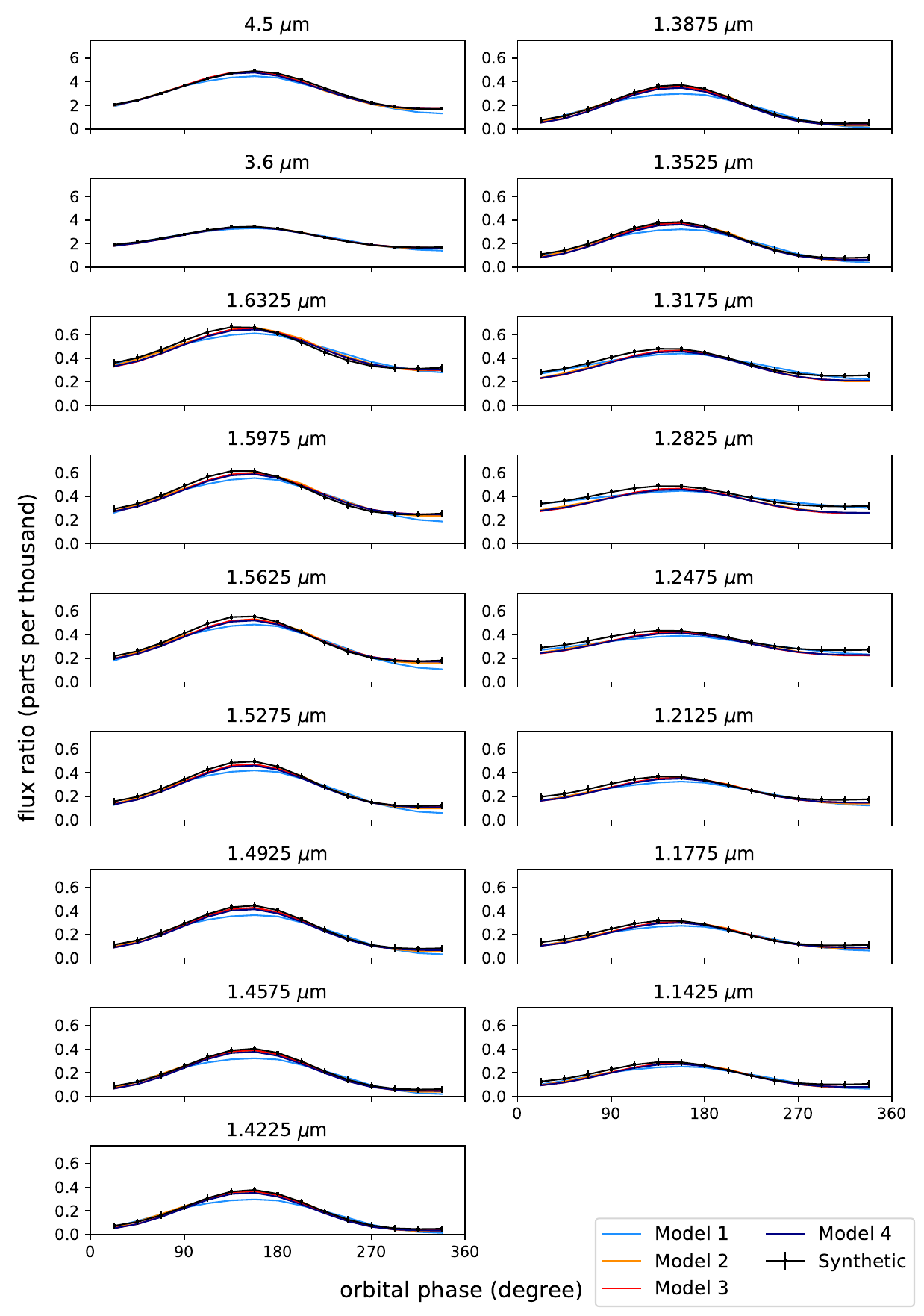}
    \caption{Results from the retrievals of synthetic phase curves generated from the WASP-43b GCM. 
    Here we plot the best-fit model phase curves calculated from the posterior medians and compared them to the synthetic data. 
    The synthetic data is shown with the measurement uncertainties of \protect\cite{stevenson_spitzer_2017}. 
    The models are described in section \ref{sec:simple_T_models}. }
    \label{fig:gcm_compare_phase_curves}
\end{figure}
\begin{figure}
    \includegraphics[width=\columnwidth]{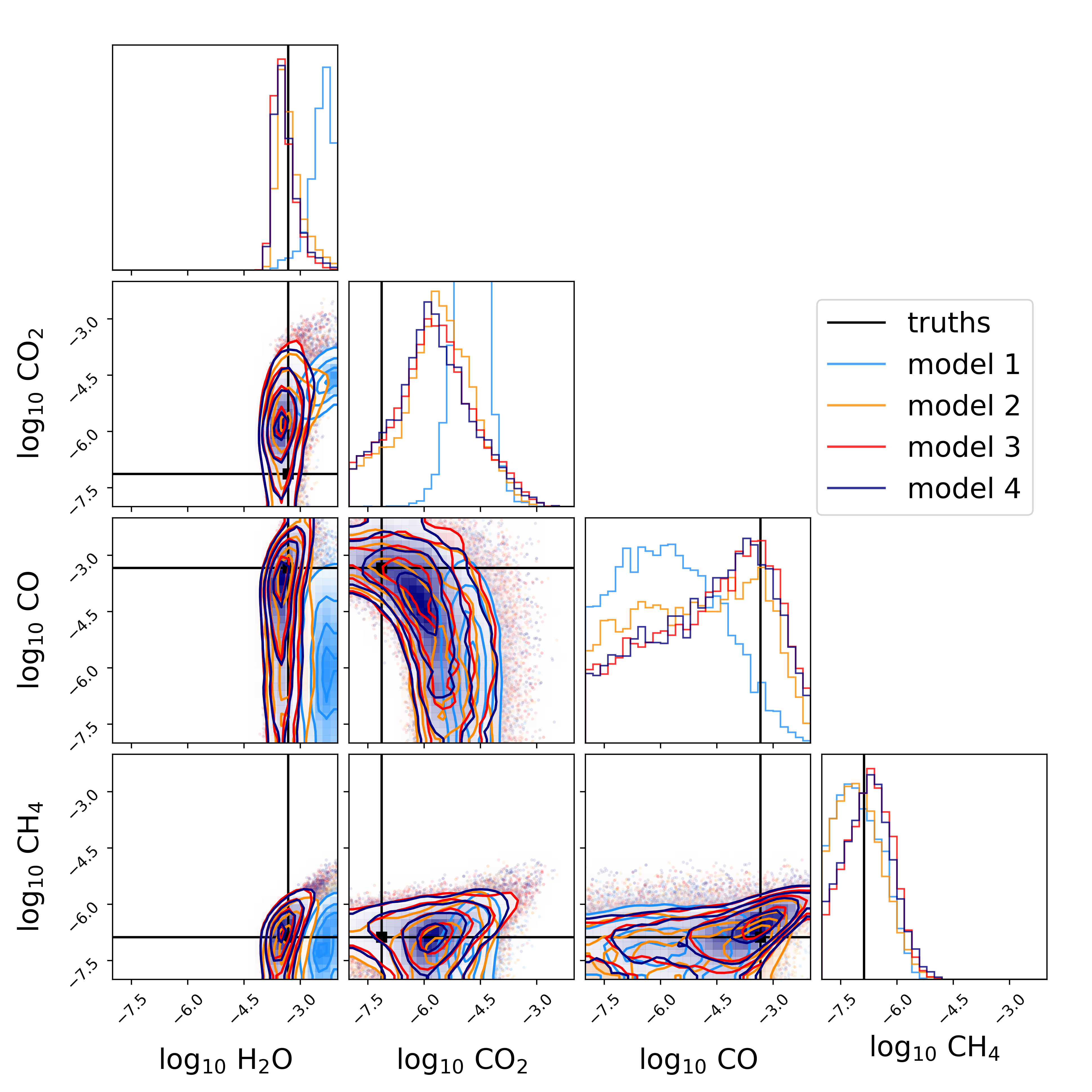}
    \caption{Results from the retrievals of synthetic phase curves generated from the WASP-43b GCM. 
    Here we plot the posterior distributions of the retrieved gas VMRs using different retrieval schemes, and we mark the abundance used to simulated the synthetic data  with black lines (`truths').}
    \label{fig:gcm_triangle}
\end{figure}
\begin{figure*}
    \includegraphics[scale=0.65]{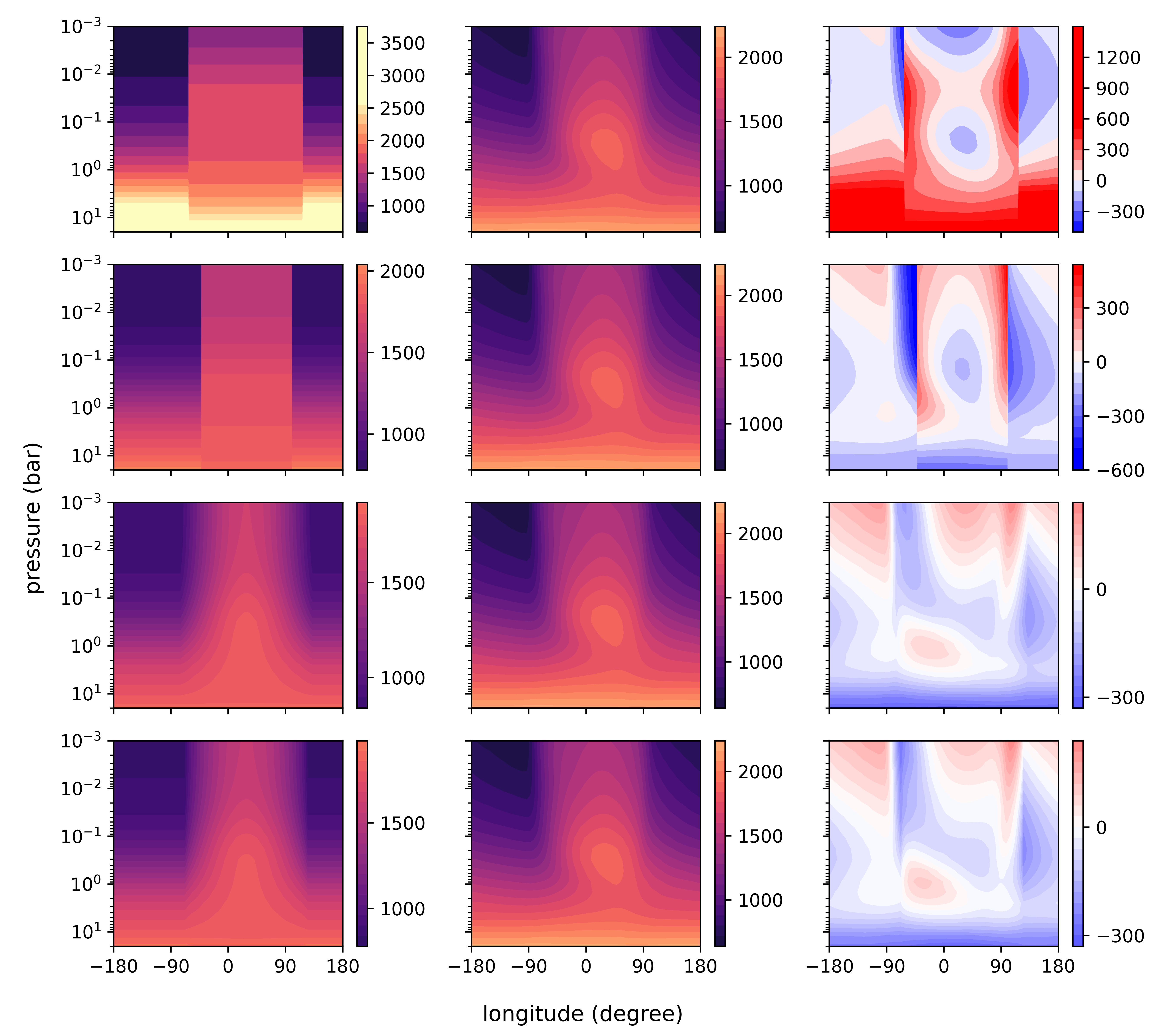}
    \caption{Results from the retrievals of synthetic phase curves generated from the WASP-43b GCM. 
    The rows from top to bottom correspond to model 1, model 2, model 3, and model 4, respectively.
    Left column: retrieved temperature structures, which are calculated using the median parameters of the posterior distributions. 
    Middle column: latitudinally averaged TP profile of the GCM from 0 to 45 degree latitude, using cos(latitude) as the weight. 
    Right column: difference between right and middle columns.}
    \label{fig:gcm_tp}
\end{figure*}
        
We validate our retrieval schemes with synthetic data simulated from the GCM-based model of WASP-43b described in section \ref{sec:gcm_intro}. 
Each retrieval scheme is identified with one of the temperature models described in section \ref{sec:simple_T_models}, while all other aspects of the retrieval schemes are identical, so we refer to each retrieval scheme by the temperature model used. 
The synthetic phase curves are simulated at the same wavelengths as those presented in \cite{stevenson_spitzer_2017}, and we use their measurement uncertainties to set the uncertainties of the synthetic data. 
We do not add random noise to the synthetic phase curves.
We assess the retrieval schemes on three criteria: (1) the goodness of fit to the synthetic phase curves; (2) the accuracy of the retrieved chemical abundance; (3) the goodness of fit of the retrieved thermal structure to the latitudinally averaged GCM thermal structure in Fig. \ref{fig:lat_mean_TP}. 
Overall, apart from model 1, all other models can fit the synthetic phase curves within measurement uncertainties at almost all wavelengths and all orbital phases, and can accurately constrain the abundance of H$_2$O and CH$_4$, as well as retrieve the latitudinally averaged thermal structure of the GCM. 

We plot the phase curves generated from the medians of  the posterior distributions of the model parameters in Fig. \ref{fig:gcm_compare_phase_curves}. 
It is clear that model 1, where the dayside\footnote{We reiterate that the `dayside' in our models denotes the region of the atmosphere modelled by the dayside profile and does not necessarily coincide with the permanently illuminated `physical dayside'.} width is fixed at 180$^{\circ}$, gives the worst spectral fit to the synthetic data. 
The phase curve amplitudes retrieved by model 1 are too small at most wavelengths, whereas the other models mostly retrieve the correct phase curve amplitudes. 
It can be seen in Fig. \ref{fig:gcm_tp} that model 1 is not flexible enough to approximate the thermal structure of the GCM, which has a hot region significantly narrower than 180 degree in longitude. 
This also leads model 1 to retrieve a biased high H$_2$O abundance, as the model tries to match the amplitudes of the synthetic phase curves by pushing the photosphere higher, where the day/night flux contrast is larger. 
We thus demonstrate that the dayside area fraction is an important parameter in phase curve retrievals, and the exclusion of its implementation can lead to significant biases in retrieved molecular abundance. 
        
We plot the posterior distributions of gas VMRs using different retrieval schemes in Fig. \ref{fig:gcm_triangle}. 
As mentioned previously, model 1 retrieves biased H$_2$O abundance due to the inflexibility of the temperature parameterisation, namely the fixed dayside fraction. 
The other models produce precise and accurate constraints on H$_2$O, as well as accurate upper bounds on CH$_4$. 
None of the models can constrain CO and CO$_2$ from the data, as  CO and CO$_2$ have weak opacities in the \textit{HST}/WFC3 wavelengths and their retrieved abundance are mainly driven by the two \textit{Spitzer} wavelengths.
Hence, their abundance are more susceptible to degeneracy with the thermal structure and are therefore poorly constrained. 

We plot the retrieved thermal structures, calculated with the median parameters of the posterior distributions, in Fig. \ref{fig:gcm_tp}. 
We compare the retrieved thermal structures to the appropriate latitudinally-averaged TP structure of the GCM, since we have shown that the appropriately averaged GCM TP structure can reproduce the synthetic phase curves generated directly from the GCM. 
As described in \ref{sec:lat_mean_tp}, the average is between 0 and 45 degree latitude and with cos(latitude) as the weight.  
It is now clear that model 1 performs badly because the width of the hot region in the GCM is significantly narrower than the dayside width prescribed by model 1. 
Model 2 can accurately approximate the typical dayside and nightside TP profiles; however, since model 2 contains a discontinuity at the dayside/nightside boundary, there are large jumps in temperature on isobars around the day/night boundary. 
Hence, the fits at those regions deviate significantly from the GCM. 
Model 3 and 4, by virtue of being continuous models, avoids this problem and can approximate the latitudinally averaged GCM structure to well within $\pm$ 300 K at most pressures and longitudes. 
However, both models perform relatively poorly in the deep atmosphere as the retrieved deep atmosphere temperatures are too low (right column, Fig. \ref{fig:gcm_tp}). 
This leads to incorrectly retrieved phase curve amplitudes at the wavelengths with the deepest photospheres.
By computing the transmission weighting function in Fig. \ref{fig:cwf}, we find the three channels at 1.2475 $\upmu$m, 1.2825$\upmu$m, 1.3175$\upmu$m are sensitive mainly to the deep atmosphere at close to 10 bar, whereas most other channels are sensitive to lower pressure levels. 
In Fig. \ref{fig:gcm_compare_phase_curves}, we can see that the phase curve fits at these three channels by model 3 and model 4 have lower flux than the synthetic phase curves.
The retrieved deep atmosphere temperature is biased because there are more data points constraining the TP profile at lower pressure levels, and the TP profile used cannot satisfy all constraints equally well.
We expect that the biased deep atmosphere temperature can be resolved by using a more sophisticated TP profile. 
However, this issue does not lead to significantly biased retrieved abundance and our precision is still in line with past studies.  

\section{Application to Real Phase Curves}

\label{sec:observation_test}
    \begin{figure}
        \includegraphics[width=\columnwidth]{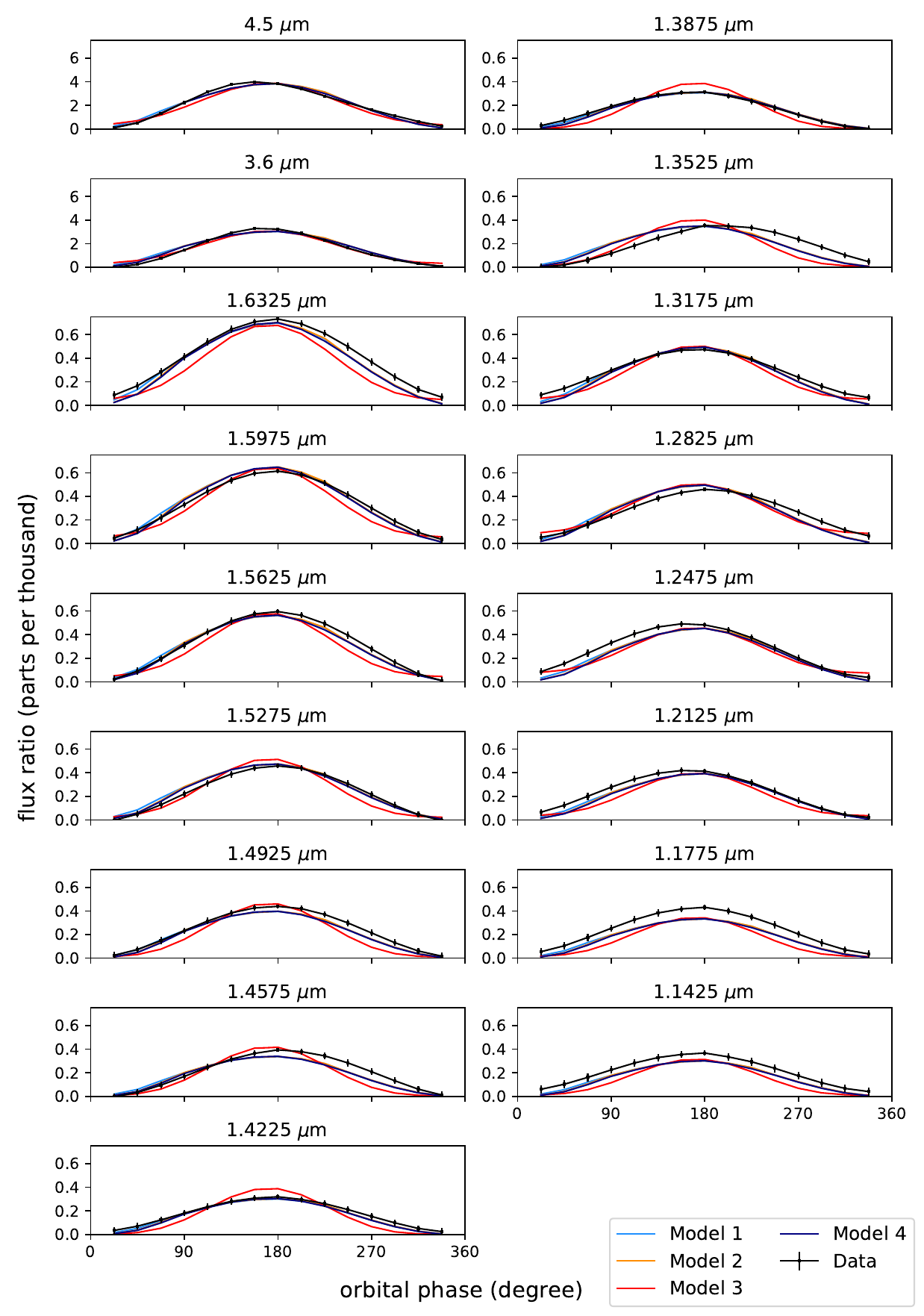}
        \caption{Retrieval results of the real phase curves. The retrieved phase curves are calculated from the posterior medians and compared to the real data. The models are described in \ref{sec:simple_T_models}.}
        \label{fig:real_compare_phase_curves}
    \end{figure}
    \begin{figure}
        \includegraphics[width=\columnwidth]{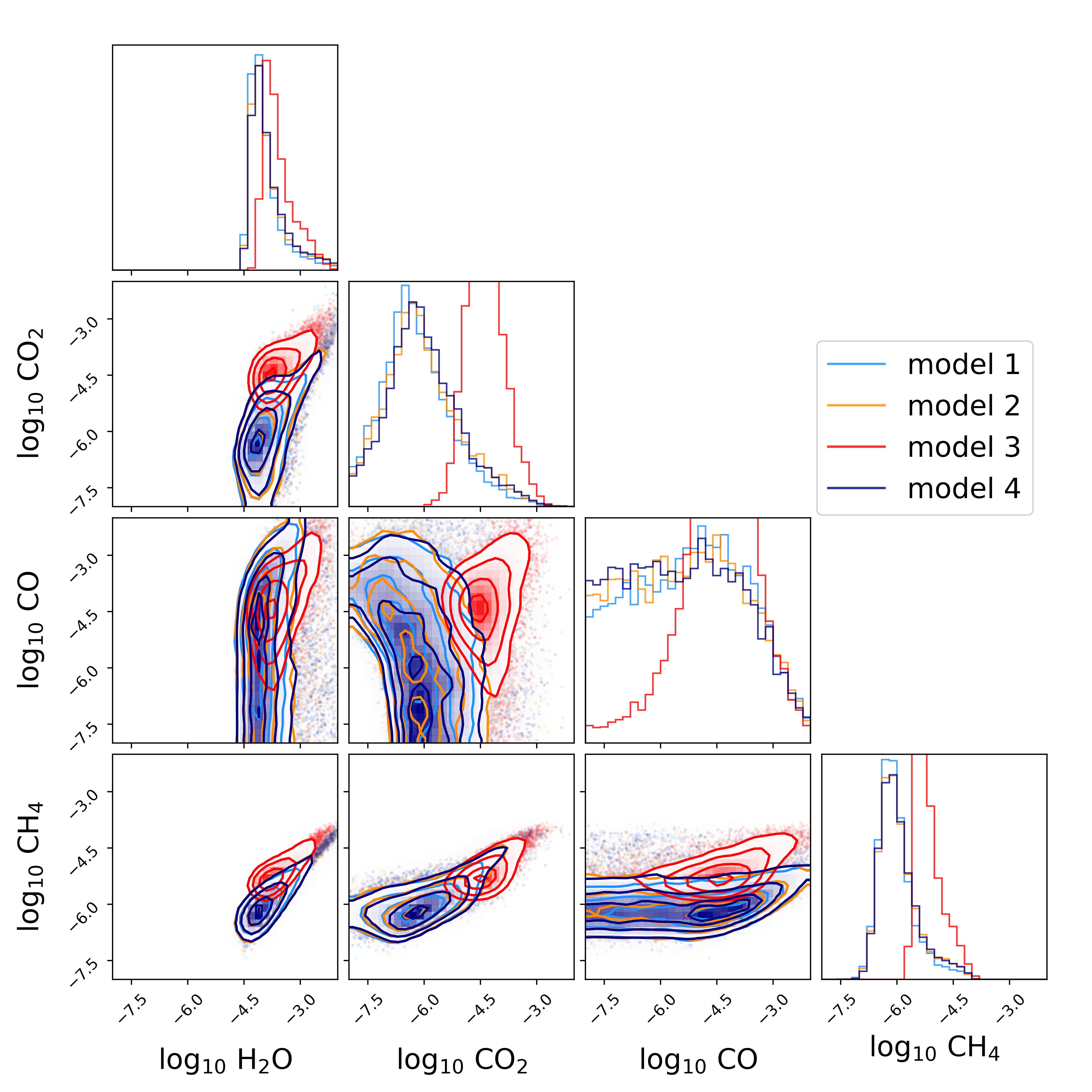}
        \caption{Retrieval results of the real phase curves. Posterior distributions of the retrieved gas VMRs.}
        \label{fig:real_triangle}
    \end{figure}
    \begin{figure*}
        \includegraphics[scale=0.65]{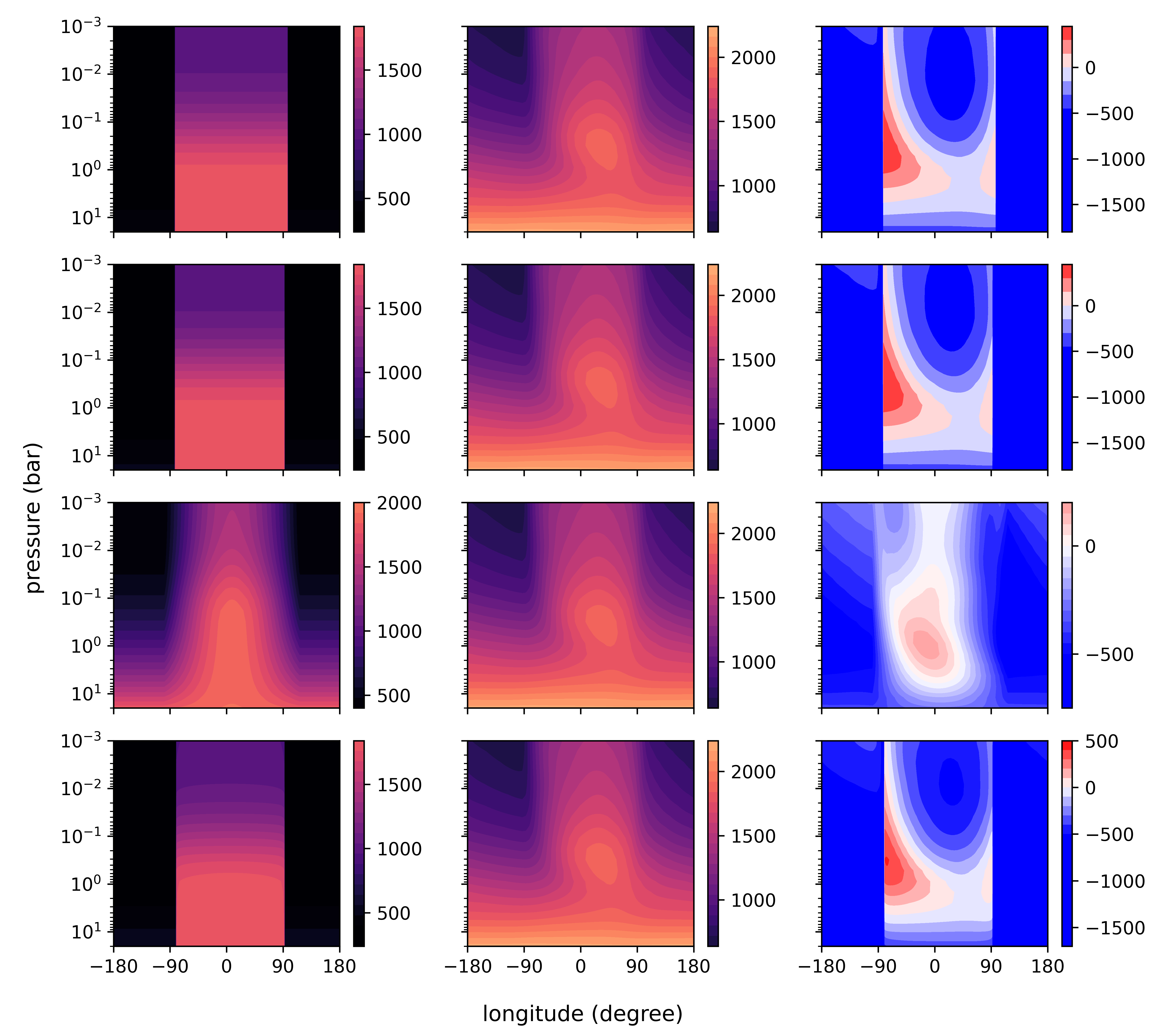}
        \caption{Retrieval results of the real phase curves. 
        The rows from top to bottom correspond to model 1, model 2, model 3, and model 4, respectively.
        Left column: retrieved temperature structures, which are calculated using the median parameters of the posterior distribution. 
        Middle column: latitudinally averaged TP profile of the GCM from 0 to 45 degree latitude, using cos(latitude) as the weight. Right column: difference between right and middle columns.}
        \label{fig:real_tp}
    \end{figure*}
    
In section \ref{sec:gcm_test}, we test the performance of our 2D retrieval schemes against synthetic data. 
We find that we can accurately constrain the abundance of H$_2$O and CH$_4$ from synthetic data simulated from a GCM, provided that the temperature model is flexible enough to approximate the thermal structure of the GCM atmosphere. 
We now apply the retrieval schemes to the observed \textit{HST}/WFC3 and \textit{Spitzer}/IRAC phase curves of \cite{stevenson_spitzer_2017}. 
We again look at the spectral fits and the retrieved chemical abundance and thermal structure. 

We plot the phase curves generated from the medians of  the posterior distributions of model parameters in Fig. \ref{fig:real_compare_phase_curves}. 
The model fits to the real data are worse than the model fits to the synthetic data, partly because we do not add random noise to our synthetic data. 
Furthermore, there are two interesting points of comparison with the fits to synthetic data. 
First, in the case of real data, model 1 can fit the phase curves almost as well as model 2, whereas in the case of synthetic data, model 1 fits the phase curves markedly worse than model 2. 
The reason that model 1 cannot fit the synthetic data is because the GCM used to generate the data has a hot region that is significantly narrower in longitudinal extent than the width of dayside region in model 1, which is fixed at 180$^{\circ}$. 
However, the thermal structure retrieved from the real data by both models is consistent with a hot dayside region that spans approximately 180$^{\circ}$ in longitude. This is the reason why model 1 could perform as well as model 2 on the real data, while failing to do so on the synthetic data.
Second, model 3 produces the poorest fits to the real data, and often produces phase curve maxima that are too large while simultaneously under-produce the amplitudes of the intermediate phases. 
Since model 3 prescribes that temperatures on isobars vary sinusoidally with longitude on the dayside, the misfits suggest that temperature must vary less strongly with longitude on isobars. 
This is confirmed by the retrieved thermal structure of model 4. 

We plot the posterior distributions of gas VMRs using different retrieval schemes in Fig. \ref{fig:real_triangle}. 
The retrieved H$_2$O abundances are consistent across models. We take model 4 as our fiducial model, which gives a constraint of  $5.6\times10^{-5}$--$4.0\times10^{-4}$ at $1\sigma$. 
As with the synthetic data, we cannot constrain the abundance of CO and CO$_2$, but we can place an upper bound on the VMR of CH$_4$ at $\sim$$10^{-6}$. We plot the retrieved thermal structures, calculated with the median parameters of the posterior distributions, in Fig. \ref{fig:real_tp}. 
The retrieved dayside temperatures of model 4 suggest that the dayside thermal structure of WASP-43b is relatively homogeneous, meaning that temperature does not vary strongly as a function of longitude on isobars. On the other hand, the retrieved temperatures on the nightside are extremely cold, which is likely due to thick cloud coverage that lifts the photosphere to lower pressure levels. We discuss the results of our retrievals in the next section.

\section{Discussions}
\label{sec:discussions}   
We compare our results to previous retrieval studies of WASP-43b, discuss the effects of nightside clouds, and detail the limitations of our retrieval model in this section.

\subsection{Comparison with previous retrievals}

\begin{table*}
\centering
\caption{Summary of previous retrieval studies of WASP-43b. The emission data analysed by \protect\cite{kreidberg_precise_2014} refer to the \textit{HST}/WFC3 secondary eclipse data as presented in \protect\cite{kreidberg_precise_2014} and the \textit{Spitzer} secondary eclipse data from \protect\cite{blecic_spitzer_2014}. The transmission data refer to the \textit{HST}/WFC3 primary transit data as presented in \protect\cite{kreidberg_precise_2014}. The phase curves refer to the \textit{HST}/WFC3 and \textit{Spitzer} phase curves as presented in \protect\cite{stevenson_spitzer_2017}. We note the constraints on the abundance of H$_2$O at 1$\sigma$ for the studies that publish such a result. For the studies that analysed phase curves, we note the retrieval methods.}
\label{tab:wasp_43b_retrieval}
\begin{tabular}{lccc}
\hline
Reference & Data  & H$_2$O 1$\sigma$ Range & Notes \\
\hline
\cite{kreidberg_precise_2014} & transmission & $3.3\times10^{-5}$--$1.4\times10^{-3}$  \\ 
\cite{kreidberg_precise_2014} & emission & $3.1\times10^{-4}$--$4.4\times10^{-3}$ & \\
 \cite{kreidberg_precise_2014} &  transmission + emission &  $2.4\times10^{-4}$--$2.1\times10^{-3}$ & \\
 \cite{stevenson_spitzer_2017} & phase curves, nightside & $2.5\times10^{-5}$--$1.1\times10^{-4}$  & phase-by-phase \\
 \cite{stevenson_spitzer_2017} & phase curves, dayside & $1.4\times10^{-4}$--$6.1\times10^{-4}$  & phase-by-phase \\
 \cite{irwin_25d_2020} & phase curves & $2\times10^{-4}$--$1\times10^{-3}$ &  `2.5D model'\\
 \cite{feng_2d_2020} & phase curves & $1.1\times10^{-4}$--$3.9\times10^{-3}$ & 2D model \\
 \cite{changeat_exploration_2021} & phase curves + transmission &  & `1.5D model' \\
 \cite{chubb_exoplanet_2022} & phase curves + transmission &   & 3D model \\ 
 This work & phase curves & $5.6\times10^{-5}$--$4.0\times10^{-4}$ & 2D model \\
\hline
\end{tabular}
\end{table*}

We present a summary of past retrieval studies of WASP-43b in Table \ref{tab:wasp_43b_retrieval}. 
We focus on the studies which analyse the \textit{HST} transmission, secondary eclipse and phase curve data \cite[GO Program 13467, PI: Jacob Bean, ][]{kreidberg_precise_2014}, \textit{Spitzer} secondary eclipse data \cite[Program ID 70084, ][]{blecic_implications_2017}, and \textit{Spitzer} phase curve data \cite[Programs 10169 and 11001, PI: Kevin Stevenson, ][]{stevenson_spitzer_2017}. 
We present the constraints on the abundance of H$_2$O at 1$\sigma$ for the studies that publish such a result. 
While all of the H$_2$O abundance constraints overlap, our constraint is on the lower end compared to past studies. Additionally, we find two points of discussion. 

Firstly, we find that multiple studies support the hypothesis that WASP-43b has no optically-thick clouds on the dayside.
\cite{kreidberg_precise_2014} analyse the transmission spectrum from \textit{HST}/WFC3, and find that the day-night terminator of WASP-43b contains no significant clouds at the pressure levels probed by transmission spectroscopy. 
Since the dayside is hotter than the terminator region and likely has comparable chemical inventory, this suggests that the dayside may be free from significant cloud coverage as well. 
More directly, \cite{fraine_dark_2021} find a very low dayside geometric albedo ($<$0.06) using \textit{HST} WFC3/UVIS secondary eclipse data in the optical wavelengths, and report a non-detection of clouds on the dayside at P $>$ 1 bar. 
Furthermore, \cite{stevenson_thermal_2014} estimate the Bond albedo of WASP-43b to be $0.18^{+0.07}_{-0.12}$ by computing the day-and night-side bolometric fluxes from the model spectra retrieved from the \textit{HST}/WFC3 phase curves.
These findings provide justifications for our cloud-free retrieval scheme on the dayside, in accordance with the prediction from modelling work that a cloud-free hot spot should dominate the dayside of hot Jupiters  \citep{parmentier_transitions_2016}. 

Secondly, we find that while some studies find variable H$_2$O abundance as a function of longitude, this might be the result of the 1D phase-by-phase approach used by these studies. 
For example, \cite{stevenson_spitzer_2017} analyse the same \textit{HST}/WFC3 and \textit{Spitzer} phase curves as in this work,  and apply a 1D phase-by-phase retrieval approach, where the spectrum at each orbital phase is retrieved independently and assuming uniform abundance and TP structure for each phase. 
They find the H$_2$O abundance varies between the dayside and the nightside phases, and give a constraint of $2.5\times10^{-5}$--$1.1\times10^{-4}$ at $1\sigma$ on H$_2$O for the nightside and a constraint of $1.4\times10^{-4}$--$6.1\times10^{-4}$ at $1\sigma$ for the dayside.
However, the 1D retrieval approach, where uniform atmospheric condition is assumed for the visible atmosphere under observation, is now known to give biased results in the analysis of disc-averaged hot Jupiter spectra \citep{blecic_implications_2017, taylor_understanding_2020}. 
Furthermore, the phase-by-phase approach is not geometrically self-consistent and under-utilises the constraints from the fact that hemispheres observed at neighbouring phases overlap \citep{irwin_25d_2020}. 
In Fig. \ref{fig:1D_vmr_synthetic} in the appendix, we plot our phase-by-phase retrieval results of the synthetic data, and show that the retrieved H$_2$O abundance varies as a function of orbital phase even though the true abundance is uniform across the planet in the GCM used to simulate the synthetic data. 
In Fig. \ref{fig:1D_vmr_real}, we plot our phase-by-phase retrieval results of the observed \textit{HST}/WFC3 and \textit{Spitzer} phase curves, and find a similar result that the retrieved H$_2$O abundance is higher on the dayside than on the nightside as \cite{stevenson_spitzer_2017}. 



\subsection{The influence of nightside clouds}
By comparing the synthetic phase curves simulated from the cloud-free WASP-43b GCM with the observed phase curves, we see that the GCM phase curves under-predict the phase curve amplitudes and over-predict the phase curve maximum offsets. 
The mismatch in phase curve offsets suggest that the strength of heat circulation is weaker on WASP-43b than predicted by the GCM, and the low nightside brightness temperatures further suggest significant nightside cloud coverage.
According to \cite{parmentier_cloudy_2020}, when nightside clouds are present, the day-to-night heat transport becomes extremely inefficient, and the nightside photosphere is lifted to higher altitude. 
This could explain the low nightside temperatures and small phase curve offsets observed on WASP-43b. 
     
\subsection{Limitations and future work}
We have introduced a new 2D retrieval scheme (model 4), where the atmospheric temperature is parameterised by equation (\ref{eq:model4}). 
We now discuss the limitations of our retrieval scheme and directions for future work.

\subsubsection{Aerosol model} 
We do not explicitly model the effects of clouds in our retrieval scheme. Past studies \citep[e.g.,][]{burningham_retrieval_2017, molliere_retrieving_2020} have shown that flexible TP profiles can mimic the spectral contribution of clouds in low-resolution spectroscopy.
We have assumed clouds with uniform-with-wavelength spectral features in the observed wavelengths are fully degenerate with thermal structure and chemical abundance. 
Disentangling this degeneracy is beyond the scope of this work. 
We recognise that the lack of cloud parameterisation is likely the most significant source of error in our retrieved atmospheric properties, though we expect the retrieved dayside properties are reliable as both transmission spectroscopy and broadband emission observation of WASP-43b find no evidence of clouds on WASP-43b \citep{kreidberg_precise_2014, fraine_dark_2021}.  
Recent studies have shown that clouds play an important role in shaping the phase curves of hot Jupiters when they are present on the nightside \citep{parmentier_cloudy_2020,roman_clouds_2021}, and we plan to include aerosols in our retrieval scheme and validate such scheme with cloudy GCMs in future work. 

\subsubsection{Temperature model}
\label{Tmodel_limitations}
Our atmospheric temperature model is strongly parameterised to keep the retrieval timescale tractable. We discuss here the limitations of the parameterisation. 

Firstly, our model is `two dimensional', meaning that temperature varies with pressure and longitude, but not with latitude. 
We then interpret the retrieved thermal structure as a latitudinally averaged thermal structure weighted towards the low latitude regions, as described in section \ref{sec:lat_mean_tp}.
\cite{irwin_25d_2020} show that the \textit{HST} + \textit{Spitzer} phase curves of WASP-43b do not allow the retrieval of the latitudinal variation of atmospheric properties, so the thermal structure of WASP-43b remains poorly constrained.
However, \textit{JWST} phase curves may allow latitudinal variation to be probed, especially when analysed in conjunction with the eclipse mapping technique \citep[e.g.,][]{rauscher_more_2018}.
The joint analysis of eclipse mapping data and phase curves will provide the most detailed constraints on the 3D structure of hot Jupiter atmospheres.
We plan to upgrade our current 2D model to include latitudinal variation in order to analyse phase curves and eclipse maps jointly in our future work, in particular the \textit{JWST}/MIRI data of WASP-43b. 

Secondly, we assume both north-south symmetry about the equator and east-west symmetry about the dayside central meridian in our temperature model. 
The assumption of north-south symmetry is based on the GCM of \cite{parmentier_transitions_2016} described in section \ref{sec:gcm_intro}, which exhibits negligible differences between the northern and southern hemispheres.
The symmetry between the northern and southern hemispheres has been seen in other hot Jupiter GCM studies as well \citep{amundsen_uk_2016, roman_clouds_2021, mendonca_angular_2020-1}. 
However, the simulations of \cite{cho_storms_2021} suggest that the atmospheres of hot Jupiter may be highly turbulent, and the atmospheric thermal structures may exhibit significant time variability that breaks the north-south symmetry.
Repeated observations of hot Jupiters can detect such time-variation. 
Since we have not validated our retrieval scheme on such turbulent atmospheres, our temperature model would require further testing when time-variation is present. 
The east-west symmetry, on the other hand, limits the flexibility of our model to capture certain thermal structures seen in GCMs, for example, temperature decreasing faster with longitude in the westward direction than in the eastward direction, or pressure-dependent hot spot shift (when the hottest hemispheres in each pressure level do not align). 
In the case of retrieving synthetic phase curves, these two effects only mildly limit our spectral fits, as shown in Fig. \ref{fig:gcm_compare_phase_curves}.
Furthermore, as seen in Fig. \ref{fig:gcm_triangle} and Fig. \ref{fig:gcm_tp}, the chemical abundance and thermal structure retrieved are not significantly biased, and are in line with literature results in terms of precision \citep[e.g.,][]{feng_2d_2020, irwin_25d_2020}.
Nevertheless, we expect such a approach would not be appropriate for data of higher quality than the ones we have analysed. 
As we plan to apply our retrieval schemes to \textit{JWST}-quality data, particularly the MIRI observation of WASP-43b, we plan to make modifications to our schemes to include such secondary structures.


\subsubsection{Chemistry model}
\label{limitation_chemistry}

We assume that the chemical abundance of gas species are constant with location and constant with pressure in our atmospheric model as we focus on the parameterisation of atmospheric temperature in this work.
The validity of this assumption depends on the gas species, the characteristics of the planetary atmosphere and the pressure range we are interested in. 
The chemical and dynamical modelling work of \cite{cooper_dynamics_2006} and \cite{agundez_pseudo_2014} suggest that for H$_2$O and CO, which are predicted to be the most abundant spectrally active molecules on hot Jupiters in the pressure ranges probed by low-resolution spectroscopy ($\sim$10-$10^{-3}$bar), the constant abundance assumption is valid as atmospheric circulation effectively homogenises their abundance. 
The assumption holds less well for CO$_2$, though we do not expect this to be a significant source of error, as the variation in CO$_2$ abundance in the models of \cite{agundez_pseudo_2014} is less than one order of magnitude.
The case of CH$_4$ is most problematic: although its abundance shows negligible variation with longitude, the vertical variation is significant \citep{agundez_pseudo_2014}.
More recently, \cite{baeyens_grid_2021} calculated a grid of pseudo-2D chemistry models for hot Jupiter atmospheres, and their results enable us to directly assess the constant chemistry assumption for a WASP-43b-like atmosphere \cite[Figure 18,][]{baeyens_grid_2021}. 
Their model suggests that the variations of H$_2$O, CO, and CO$_2$ abundance are well within one-order of magnitude in the pressure range 100-$10^{-4}$bar, both with respect to longitude and with respect to pressure. 
This pressure range should more than cover the pressure range probed by low-resolution infrared emission spectroscopy of WASP-43b. 
Since typically even the best molecular abundance constraints from \textit{HST} + \textit{Spitzer} data have uncertainties around one order of magnitude, we conclude that for H$_2$O, CO, and CO$_2$ it is valid to assume constant chemistry for \textit{HST} + \textit{Spitzer} data based on the cited modelling work.
As for CH$_4$, the model of \cite{baeyens_grid_2021} suggests that for a WASP-43b like atmosphere while the CH$_4$ abundance is significant in the deep atmosphere (at pressure level greater than about 1 bar), its abundance rapidly decreases with decreasing pressure. 
If this is the case, then our retrieved upper bound of CH$_4$ abundance at around 10$^{-6}$ would not reflect the true CH$_4$ abundance of the atmosphere, which would be highly pressure-dependent.
Apart from the interplay between circulation and equilibrium chemistry, photochemistry and molecular dissociation can also affect the spatial distribution of molecular abundance. 
However, we only expect dissociation to be significant in much more strongly irradiated planets than WASP-43b, and we expect photochemical products to be insignificant in the pressure region probed by emission spectroscopy. 
For \textit{JWST} data, we expect that the constant abundance approximation could still be valid for H$_2$O, CO, and CO$_2$ when analysing emission spectroscopy of hot Jupiters, based on the current modelling work. 
However, more sophisticated parameterisation of abundance variation is necessary for CH$_4$, and for joint analysis of transmission and emission data where a large pressure range is probed.

\section{Conclusions}
\label{sec:conclusions}
We propose a novel 2D temperature parameterisation for the retrievals of hot Jupiter phase curves, which is described by equations (\ref{eq:model4}). 
The temperature model is a function of pressure and longitude, and can be used to retrieve the chemical composition and latitudinally averaged thermal structure of hot Jupiters atmospheres from phase curves. 
The model is built on two TP profiles, signifying the representative profiles for the dayside and the nightside.
In our model, the temperature is uniform on isobars on the nightside, and varies with cos$^n$(longitude/$\epsilon$) on isobars on the dayside, where $n$ and $\epsilon$ are free parameters.
Both the dayside central longitude and dayside fraction (longitudinal extent) are free parameters of our model. 
We first apply our proposed retrieval scheme, together with several other 2D models for comparison, to synthetic phase curves simulated from a cloud-free GCM of WASP-43b, representing a more realistic atmospheric model than the typically simple models used for validating retrieval schemes in the literature.
We find that the models that allow variable dayside longitudinal extent can fit synthetic \textit{HST}/WFC3 and \textit{Spitzer}/IRAC phase curves to within typical measurement uncertainties, as well as accurately and precisely constraining the water and methane abundance. 
The retrieved thermal structures using these models are good approximations to the latitudinal-average of the GCM thermal structure weighted towards the low latitude regions. 
We then apply our retrieval schemes to retrieve information from the observed phase curves of WASP-43b presented in \cite{stevenson_spitzer_2017}. 
We constrain the abundance of water to be $5.6\times10^{-5}$--$4.0\times10^{-4}$ at $1\sigma$ using model 4, as well as an upper bound on CH$_4$ at $\sim 1\times10^{-6}$. 
We find that the latitudinally averaged dayside TP structure of WASP-43b is likely to be homogeneous (meaning that temperature does not vary strongly as a function of longitude on isobars) and non-inverted. 
We expect the nightside of WASP-43b to be covered by thick clouds due to the extremely low retrieved nightside temperature, in agreement with previous studies. 
We have thus demonstrated the efficacy of our retrieval scheme, which simultaneously fits all orbital phases of a set of phase curves at a modest computation cost.  

\section*{Acknowledgements}
JKB is supported by a Science and Technology Facilities Council Ernest Rutherford Fellowship, grant number ST/T004479/1.

\section*{Data availability}
The data underlying this article will be shared on reasonable request to the corresponding author.




\bibliographystyle{mnras}
\bibliography{Paper1} 



\appendix
\section{\textsc{nemesispy}}
\label{app:nemesispy}
The radiative transfer transfer calculations, disc-averaging and temperature models are all implemented by our open-source Python software \textsc{nemesispy}, available at \url{https://github.com/Jingxuan97/nemesispy} or from the Python Package Index  at \url{https://pypi.org/project/nemesispy}. 
The package is based on the FORTRAN \textsc{NEMESIS} library \citep{irwin_nemesis_2008}, with substantial code refactoring to improve computational speed, as well as new developments, including the implementation of several 2D retrieval schemes for analysing exoplanet phase curves. 
The most computationally expensive routines are compiled to machine code at runtime using a just-in-time (JIT) compiler\footnote{\url{https://numba.pydata.org/numba-doc/latest/user/jit.html}}, so that the speed of our code is on par with compiled languages. 

\section{Comparison with phase-by-phase retrievals}
\label{app:phase_by_phase}
\begin{figure*}
\includegraphics[scale=0.70]{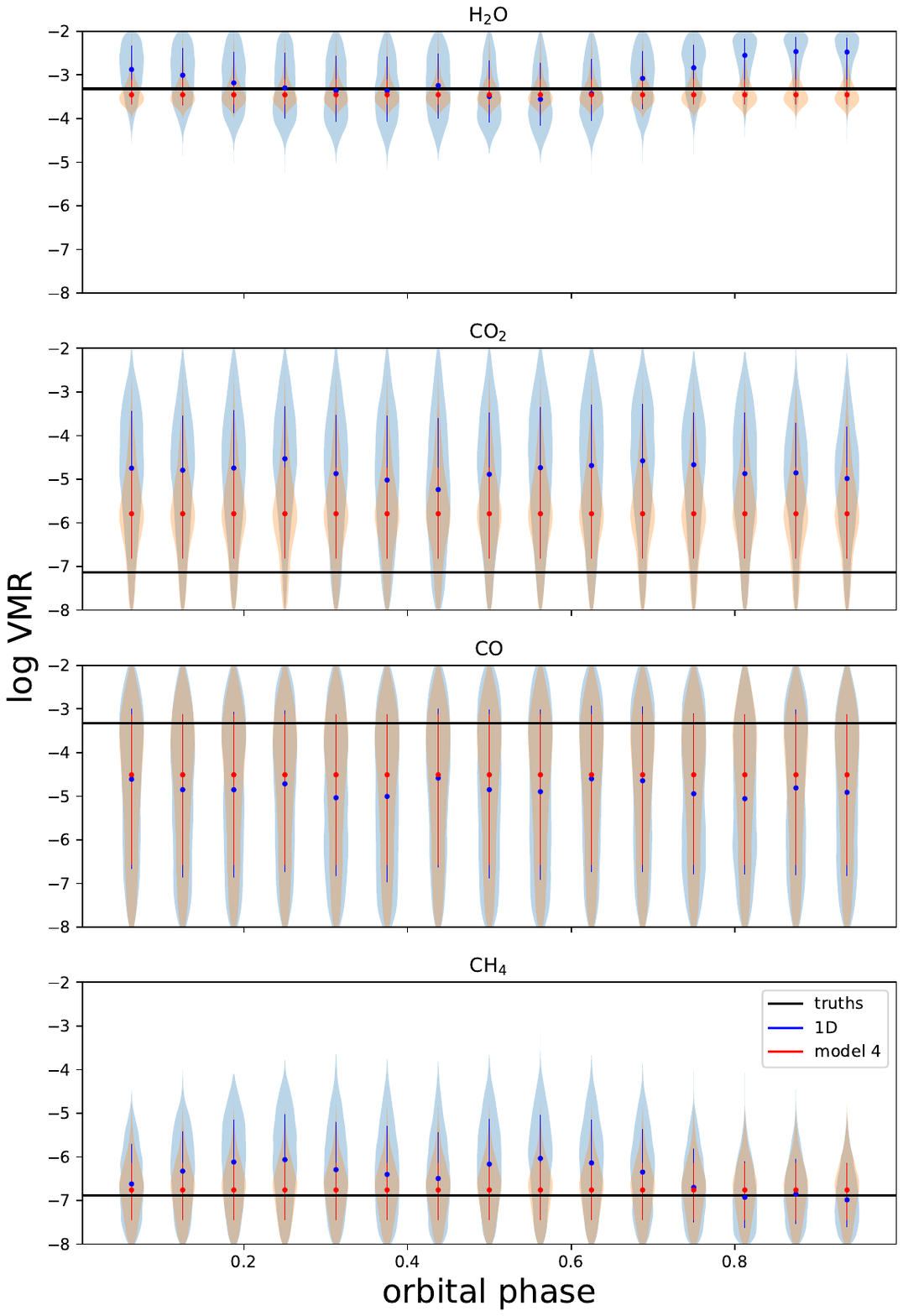}
\caption{Retrieved molecular abundance from the synthetic phase curves using the 1D phase-by-phase approach (blue), compared to the retrieved abundance using model 4 (red). The vertical lines mark the 1 $\sigma$ confidence intervals. The true abundances used to generate the data are marked by the black horizontal lines.}
\label{fig:1D_vmr_synthetic}
\end{figure*}

\begin{figure*}
\includegraphics[scale=0.70]{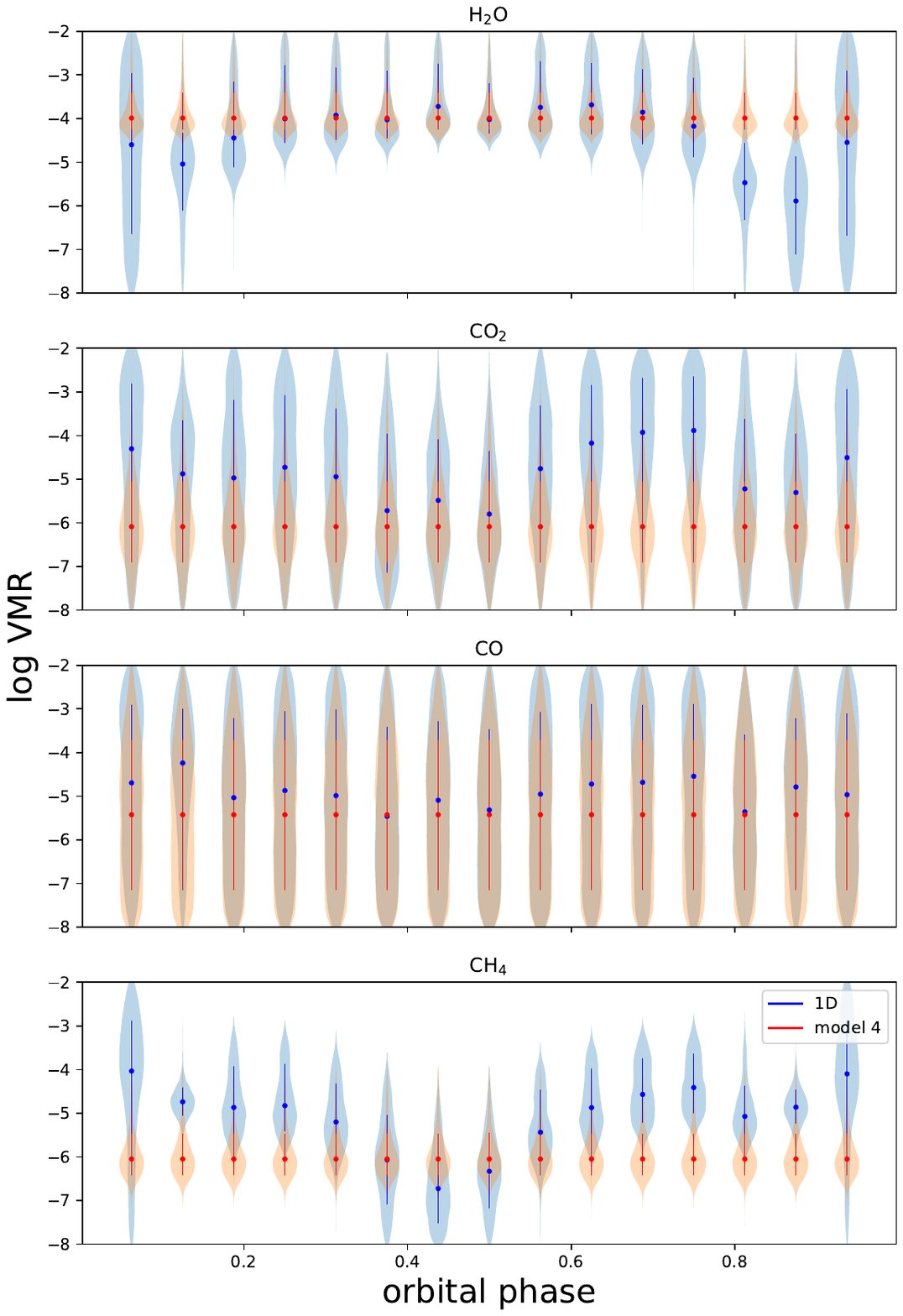}
\caption{Retrieved molecular abundance from the observed phase curves presented in \protect\cite{stevenson_spitzer_2017} using the 1D phase-by-phase approach (blue), compared to the retrieved abundance using model 4 (red). The vertical lines mark the 1 $\sigma$ confidence intervals. }
\label{fig:1D_vmr_real}
\end{figure*}

We show the retrieval results using the 1D phase-by-phase approach, where the spectrum at each orbital phase is independently analysed with a uniform TP and abundance profile. 
The TP profile used for the 1D retrievals is the same Guillot profile that we use for the 2D retrievals.
In Fig. \ref{fig:1D_vmr_synthetic}, we present the retrieved molecular abundance from the synthetic phase curves, where the truths are marked by horizontal black lines. We see that the retrieved H$_2$O abundance using the 1D phase-by-phase approach varies significantly with orbital phases, and is biased at several orbital phases. 
We echo the findings of past studies \citep{blecic_implications_2017, taylor_understanding_2020} that the 1D phase-by-phase approach could lead to significantly biased molecular abundance. 
In Fig. \ref{fig:1D_vmr_real}, we present the retrieved molecular abundance from the real phase curves presented in \cite{stevenson_spitzer_2017}. 
We find a similar trend as \cite{stevenson_spitzer_2017}, that H$_2$O abundance is higher for the dayside phases than for the nightside phases. 
We plot the abundance constraints from model 4 on the figures for comparison. 
We can see from Fig. \ref{fig:1D_vmr_synthetic} that model 4 performs markedly better than the 1D approach on synthetic phase curves both in terms of accuracy and precision in retrieved molecular abundance.

\section{Full posterior plots}
We include the full posterior distributions of all models for the retrievals of the synthetic data.
We additionally include the full posterior distribution of model 4 for the retrievals of the observed data.

\begin{figure*}
\includegraphics[scale=0.2]{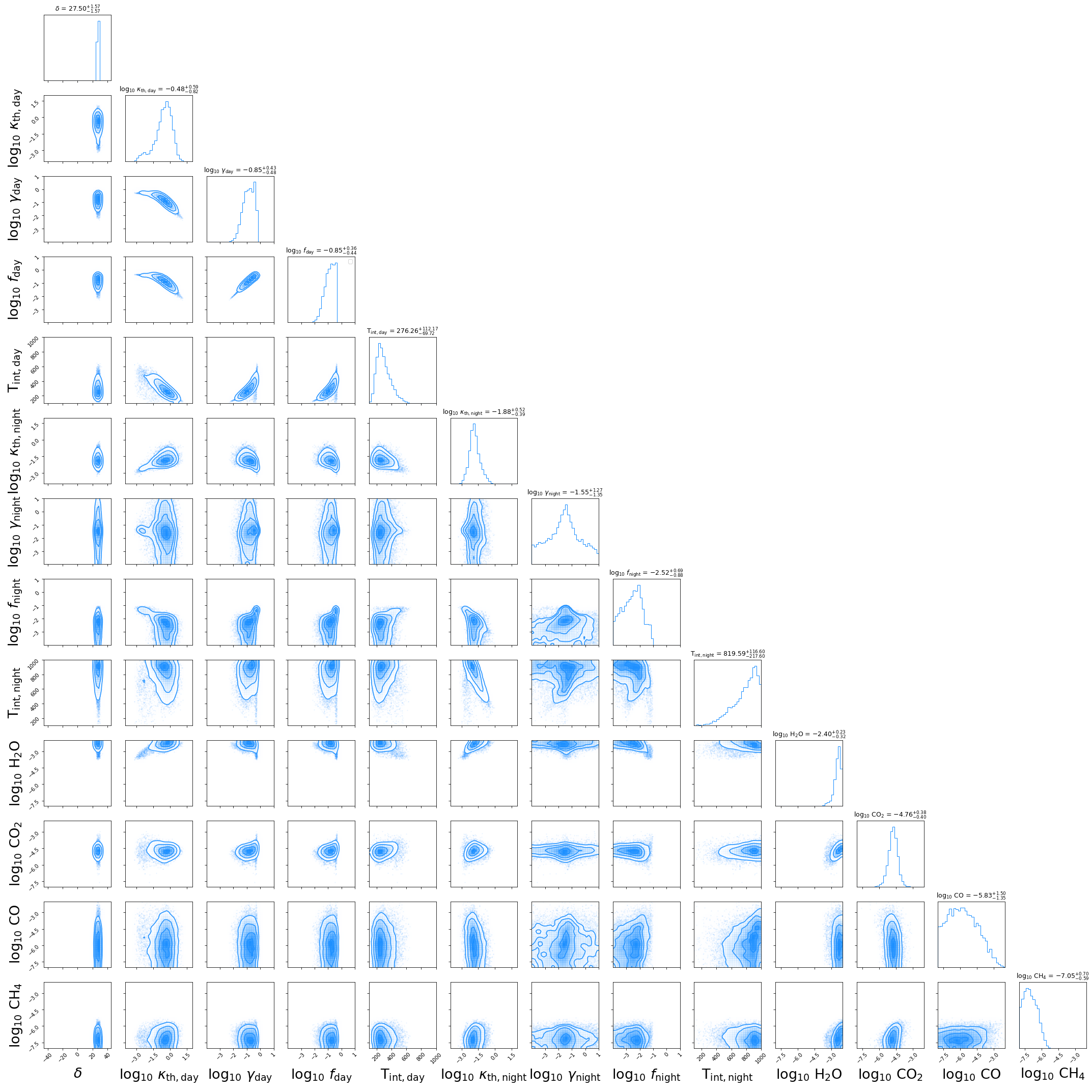}
\caption{Full posterior distribution of the model parameters of model 1 for the retrievals of the synthetic phase curves. The parameters are summarised in Table \ref{tab:priors}.}
\end{figure*}

\begin{figure*}
\includegraphics[scale=0.20]{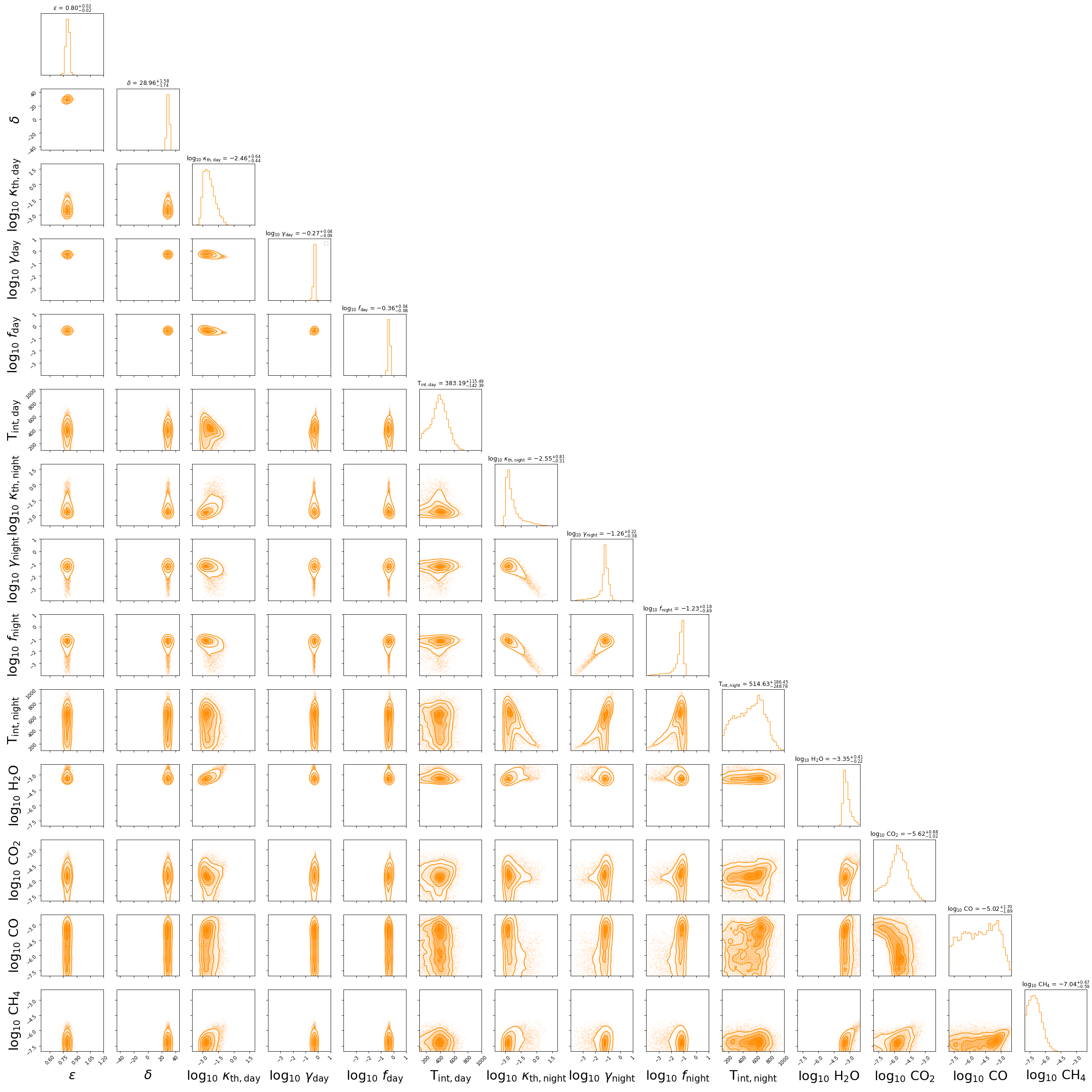}
\caption{Full posterior distribution of the model parameters of model 2 for the retrievals of the synthetic phase curves. The parameters are summarised in Table \ref{tab:priors}.}
\end{figure*}

\begin{figure*}
\includegraphics[scale=0.20]{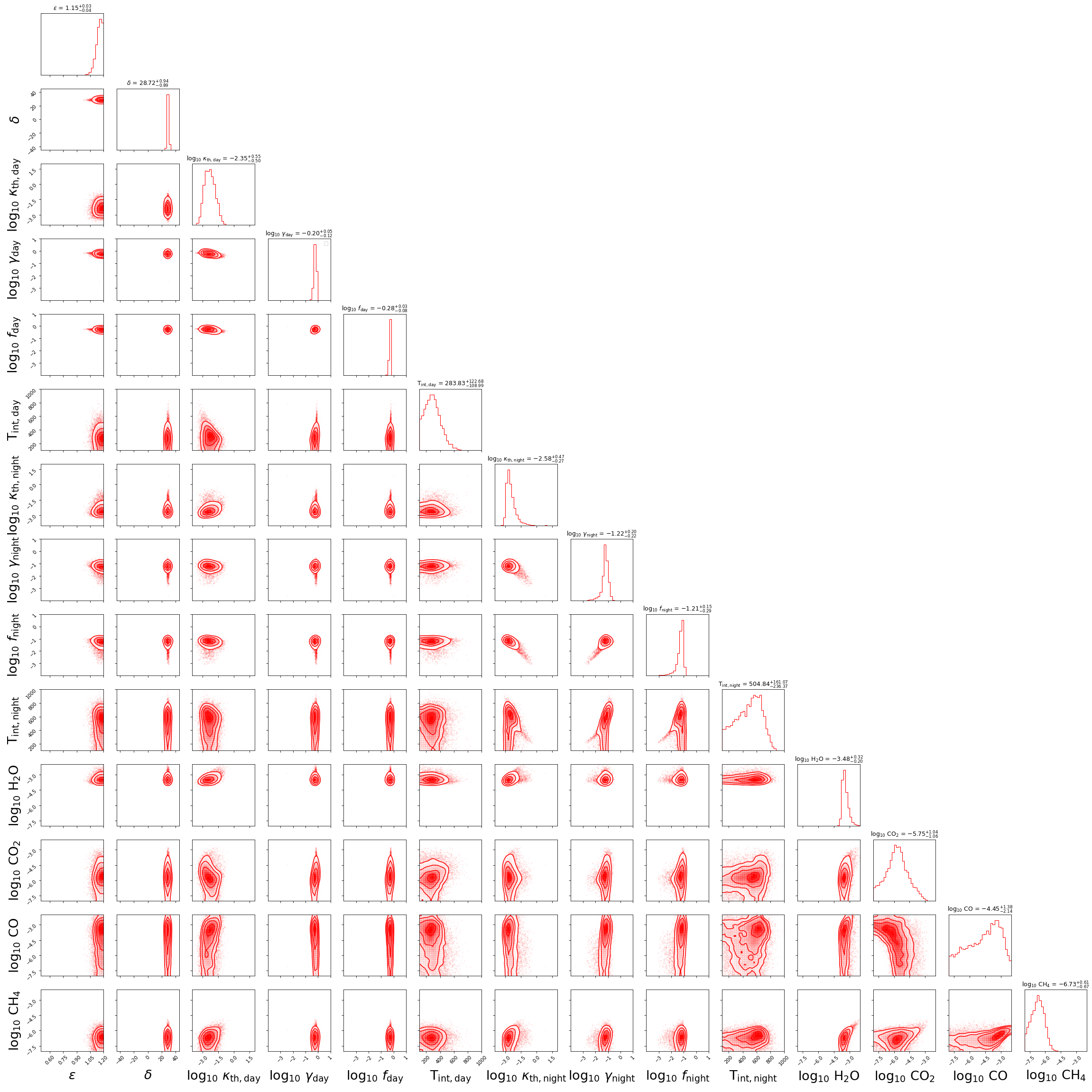}
\caption{Full posterior distribution of the model parameters of model 3 for the retrievals of the synthetic phase curves. The parameters are summarised in Table \ref{tab:priors}.}
\end{figure*}

\begin{figure*}
\includegraphics[scale=0.20]{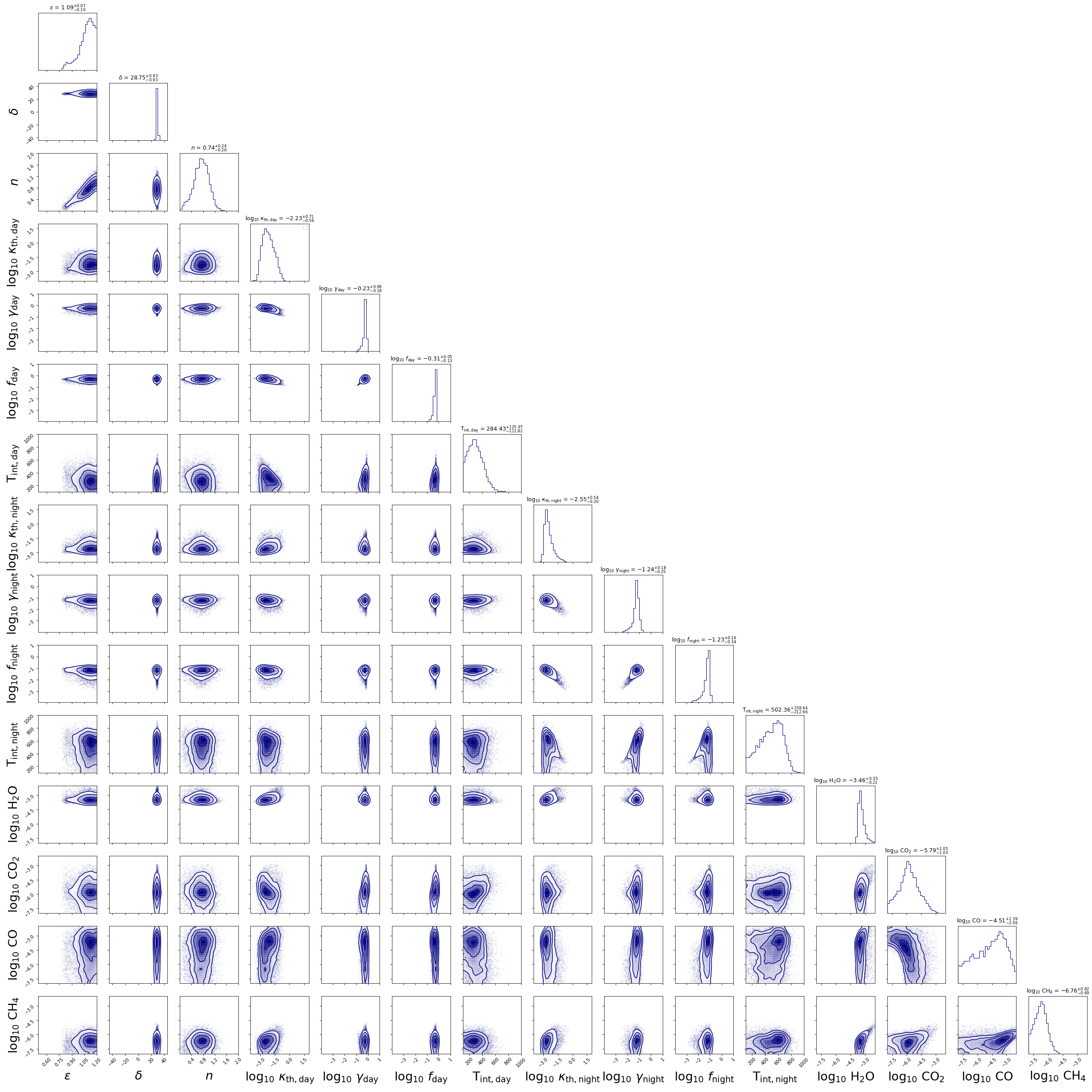}
\caption{Full posterior distribution of the model parameters of model 4 for the retrievals of the synthetic phase curves. The parameters are summarised in Table \ref{tab:priors}.}
\end{figure*}

\begin{figure*}
\includegraphics[scale=0.20]{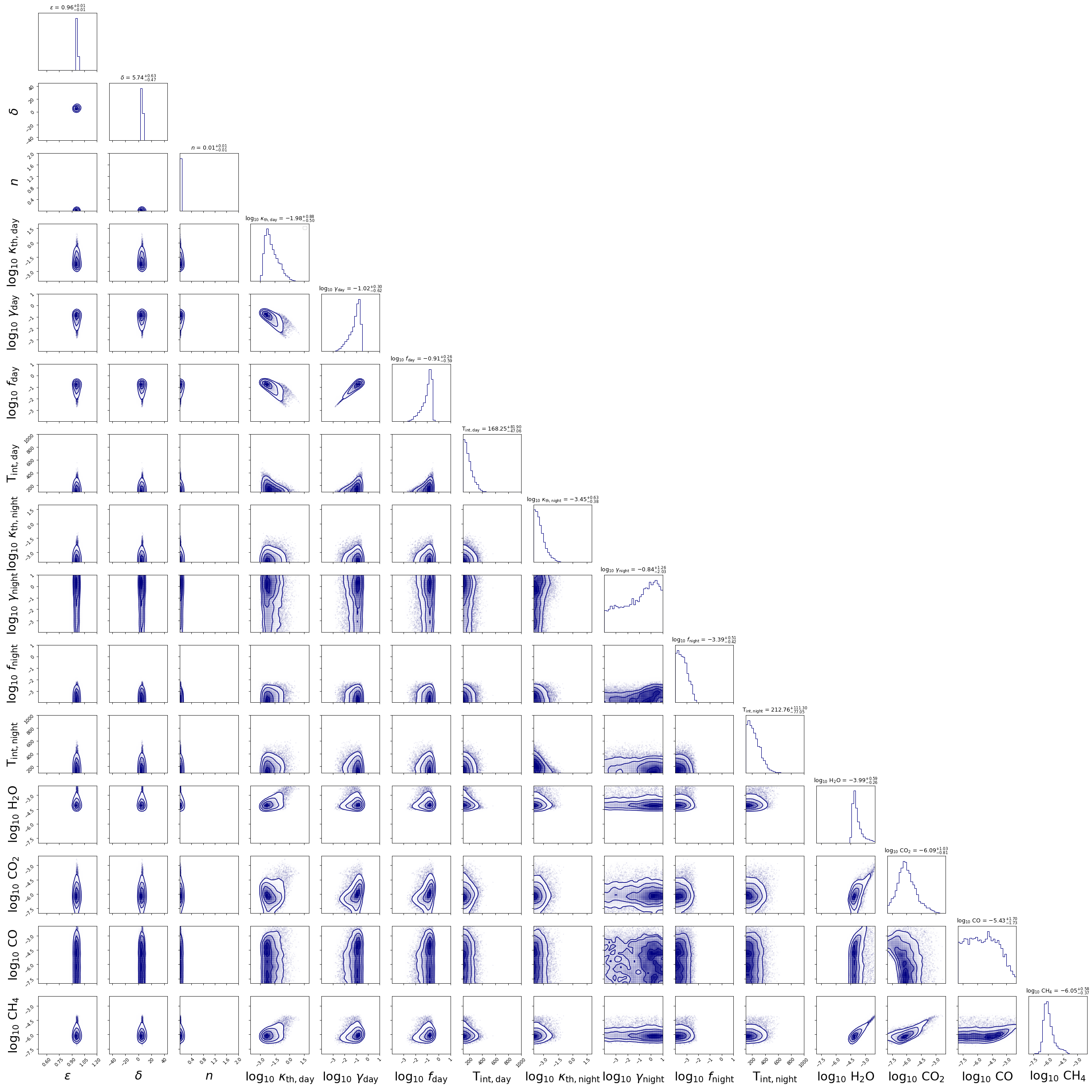}
\caption{Full posterior distribution of the model parameters of model 4 for the retrieval of the observed phase curves. The parameters are summarised in Table \ref{tab:priors}.}
\end{figure*}


\bsp	
\label{lastpage}
\end{document}